\documentclass[12pt,leqno]{article}
\usepackage{amssymb}
\usepackage[mathscr]{eucal}
\usepackage{amsmath,amssymb,latexsym,theorem,bbm,url}
\usepackage{color}
\usepackage{graphicx}
\usepackage{appendix}
\usepackage{subcaption}

\newcommand{\NN}{\mathbb{N}}

\newcommand{\RR}{\mathbb{R}}

\newcommand{\bF}{{\boldsymbol{F}}}

\newcommand{\bv}{{\boldsymbol{v}}}

\newcommand{\bzero}{{\boldsymbol{0}}}

\newcommand{\cA}{{\mathcal A}}
\newcommand{\cB}{{\mathcal B}}

\newcommand{\dd}{\mathrm{d}}

\newcommand{\EE}{\operatorname{\mathbb{E}}}
\newcommand{\DD}{\operatorname{\mathbb{D}}}
\newcommand{\PP}{\operatorname{\mathbb{P}}}

\newcommand{\rsd}{\operatorname{RSD}}

\renewcommand{\mid}{\,|\,}

\renewcommand{\leq}{\leqslant}
\renewcommand{\geq}{\geqslant}

\newcommand{\proofend}{\hfill\mbox{$\Box$}}

\numberwithin{equation}{section}

\theoremstyle{change} \theorembodyfont{\em}
\newtheorem{Lem}{Lemma.}[section]

\theorembodyfont{\rm}
\newtheorem{Rem}[Lem]{Remark.}

\newtheorem{Exe}[Lem]{Exercise.}

\begin{document}

\begin{center}
 {\bfseries\Large A stochastic approach in physics exercises\\[1mm] of mathematics education}

\vspace*{3mm}

 {\sc\large
  M\'aty\'as $\text{Barczy}^{*,\diamond}$,
  \ Imre $\text{Kocsis}^{**}$,
  \ Csaba G\'abor $\text{K\'ezi}^{**}$}

\end{center}

\vskip0.2cm

\noindent
 * HUN-REN–SZTE Analysis and Applications Research Group,
   Bolyai Institute, University of Szeged,
   Aradi v\'ertan\'uk tere 1, H--6720 Szeged, Hungary.

\noindent
 ** Department of Basic Technical Studies,
    Faculty of Engineering, University of Debrecen,
    \'Otemet\H o utca 2-4, H--4028 Debrecen, Hungary.

\noindent E-mails: barczy@math.u-szeged.hu (M. Barczy),
                   kezicsaba@eng.unideb.hu (Cs. G. K\'ezi),\\
                   kocsisi@eng.unideb.hu (I. Kocsis).

\noindent $\diamond$ Corresponding author.

\renewcommand{\thefootnote}{}

\vspace*{1mm}

\begin{abstract}
We present a method for incorporating a stochastic point of view into physics exercises of mathematics education.
The core of our method is the randomization of some inputs, the system model used does not differ
 from what we would use in the deterministic approach.
We consider exercises from the theory of projectile motion and statics.
The outputs of stochastic models are random variables, and we usually determine
  their probability distributions, expected values, variances, and relative standard deviations,
  and the probabilities of some events related to them are also calculated.
Students and teachers familiar with elementary probability theory and mechanics may
 find these exercises useful for understanding some basic concepts of stochastic mechanics.
\end{abstract}

\section{Introduction}
\label{section_intro}

Deterministic and stochastic models play equally significant roles in engineering sciences and economy.
However, in many countries (for example, in Hungary) a  deterministic approach is at the heart of education and thus of student thinking
 (in particular, at high schools).

The effect of uncontrolled input (noise) on the output of a system can be negligible,
 but it can also be very important and lead to false conclusions.
In deterministic models, we make the simplifying assumption that there is no noise.
Whether this simplification causes problems in solving a real-world problem depends on how much uncertainty there is in the system's inputs and parameters,
 and how sensitive the output is to this uncertainty.
If we do not examine this, then we cannot judge the adequacy of the solution resulting from the deterministic model.
When solving a problem, a conscious choice must be made between deterministic and stochastic models.
Didactically, the applicability of the two approaches is very different.
A stochastic approach is usually unavoidable when analyzing problems in engineering sciences, economy and financial mathematics.
The emphasis is on seeing the difference in the output (solution of the problem under study) for deterministic and random inputs.
Studying the stochastic approach is important for understanding the phenomena and processes of everyday life.
For example, we constantly evaluate risks and act according to our individual risk tolerance.
A person who is not familiar with the concept of risk does not understand what to expect from an investment
 and is unable to deal with situations that do not correspond to his or her preconcept.

There is a long tradition of the illustration of the basic ideas and calculations of probability theory in teaching
 using gambling games.
For example, we can mention the book of Schoenberg \cite{Sch} in which both standard and advanced probability topics
 are illustrated using the popular poker game of Texas Hold'em, rather than the typical balls in urns.
Johnson \cite{Joh} has chosen the gambling game of Craps for such purposes.

In this article, we present a method for incorporating a stochastic point of view into physics exercises of mathematics education.
The core of the method is the randomization of some inputs, the system model used does not differ
 from what we would use in the deterministic approach.
We take as a basis the mathematical models of some real physical phenomena and we incorporate randomness into this mathematical
 model by randomizing some of the inputs.
In our exercises, the distribution of random inputs, in most of the cases, is a uniform distribution.
This does not weaken the pedagogical goal, since using inputs with more complicated distributions would not make a difference
 in principle, it would only make the calculations more difficult.
Of course, in real life problems, where the distribution of the random inputs can be more general than uniform distributions,
 as a first step, one should approximate the distribution functions of the inputs, for example, by the help of
 empirical distribution functions.
The outputs of stochastic models are random variables, and even if the inputs are uniformly distributed,
 the outputs are usually not uniformly distributed.
We note that in Exercise \ref{Ex_proj_motion_4} an absolutely continuous random variable with a given density function also
 appears for the ratio of the rebound and landing speeds when a tennis ball is dropped on concrete from a given distance.
In the exercises, we usually determine the probability distributions, expected values, variances, and relative standard deviations
 of the outputs, and the probabilities of some events related to them are also calculated
 using the basic tools of probability theory.
In real life problems, one can usually give only approximations of the quantities in question (possibly using some software),
 since, as we mentioned above, in practice the distributions of the random inputs are approximated as well.
We emphasize that we do not develop any new mathematical model for physical phenomena, we just
 incorporate randomness into the existing models by considering random inputs having
 uniform distributions in most of the cases (for simplicity).
Having stochastic models at hand, in our exercises, we present several methods for solving probabilistic problems
 related to the random outputs of the models in question.

We demonstrate our method of randomization of some inputs by considering exercises from the theory of projectile motion and statics.
According to our knowledge, this stochastic viewpoint is not widespread for problems in projectile motion,
 and basically is not present at high schools and at physics competitions.
Nonetheless, we can mention the research paper of Montecinos et al.\ \cite{MonDavPer} where an application
 of the maximum entropy inference to the problem of a two-dimensional projectile motion was investigated.
Namely, given information about the average horizontal range over many realizations,
 they studied the problem of inferring the initial conditions, i.e., the initial angle and speed.
In statics one can find several stochastic models.
Here we only mention a recent paper of Cort\'es et al.\ \cite{CorLopNavRomRos} who carried out
 a probabilistic analysis of the deflection of a cantilever beam subjected to
 random loads via density functions.

Our reason for choosing the topics of projectile motion and statics for demonstrating our method 
 is the fact that the physics behind these two topics is relatively easy to understand. 
In Appendices \ref{appProbab} and \ref{appPhys}, we collected some of the basics of
 probability theory and mechanics that are needed to solve the problems, respectively.
We mention that, in Appendix A and in the solutions of some of the exercises,
 absolutely continuous random variables and their expected value and variance come into play.
In high schools it may be not part of the standard curriculum,
but in such a situation interested teachers can include it for
the sake of interested students.
At universities, every course on basic probability theory contains
 the topic of absolutely continuous random variables as well.
The content of Appendix \ref{appPhys} on some knowledge of mechanics is essentially contained
 in the syllabus of high school students, and hence this appendix is not so detailed.

The solutions of our exercises can be successfully understood 
 by anyone who is familiar with the content of Appendices A and B.
In particular, by high school students who undertake extension work in the mathematical sciences. 
In some classes in specialized mathematics, high school students can
 have 7 or 8 hours per week in mathematics, and the teachers may have the possibility
 to present material from probability theory that are even not included in the ''standard'' syllabus
 (such as absolutely continuous random variables). In this way, students can also practice
 the integration of some elementary functions that is part of the ''standard'' syllabus in
 specializes math classes. Further, there are so-called non-compulsory workshops, 
 study circles for interested math students, who may prepare for some competitions, 
 and at these occasions the teachers may also consider our randomized exercises. 
Concerning university students, we think that our randomized exercises are definitely understandable 
 for those university students who took a course on basic probability theory.

We take great care to demonstrate the use of the tools from probability theory,
 and hence we provide detailed solutions to the exercises.
Our stochastic perspective can help the students
 recognize the fact that many phenomena in our daily lives have a random nature,
 and therefore it can be useful to get in touch with stochastic thinking and tools
 in order to be able to understand these phenomena much better.
Our exercises are related to projectile motion and statics, and we mention that 
 STEM students may be asked to create similar exercises, for other fields of physics as well.

The paper is structured as follows.
Sections \ref{section_projectule_motion} and \ref{section_statics} are devoted to
 exercises for projectile motion and statics, respectively.
In Sections \ref{section_projectule_motion} and \ref{section_statics}, we present five and three exercises, respectively.
We close the paper with three appendixes.
In Appendices \ref{appProbab} and \ref{appPhys}, without being exhaustive, we recall some of the notations, conventions, concepts and results of probability theory
 and mechanics that are used in solving the exercises.
Appendix \ref{appAltSol} is devoted to an alternative solution for calculating some probabilities
 for part (ii) of  Exercise \ref{Ex_proj_motion_1} using the law of total expectation instead of the method of geometric probabilities.

Next, we briefly summarize what kind of tools in probability theory we use in the exercises of Sections \ref{section_projectule_motion} and \ref{section_statics}.
In Exercise \ref{Ex_proj_motion_1}, among others, we use the notions of expectation, variance and relative standard deviation of an absolutely continuous random variable,
 and we calculate the probability of an event by two methods, namely, by the method of geometric probability and by the law of total expectation, respectively.
In part (iii) of Exercise \ref{Ex_proj_motion_1}, we calculate a conditional probability as well.
In Exercise \ref{Ex_proj_motion_2}, we use the notions of expectation, variance and relative standard deviation
 of a discrete random variable, and the additivity property of probability.
In Exercise \ref{Ex_proj_motion_3}, among others, the notion of the distribution
 of discrete and absolute continuous random variables and the notions of expectations, variances and relative standard deviations of such
  random variables come into play.
In Exercise \ref{Ex_proj_motion_4}, we use the fact that a probability that an absolutely continuous random variable
 falls in a given interval can be calculated using the density function of the random variable in question,
 and we provide an application of Bayes' theorem as well.
In Exercise \ref{Ex_proj_motion_5}, we calculate the expected value of the relative error
 in some problems for free fall (i.e., for a motion of a body, where the gravity of Earth is the only force acting upon it)
 if instead of the precise value of the gravity of Earth we use an approximating random value for it.
In this exercise, the notion of the expected value of an absolutely continuous random variable is used.
In Exercises \ref{Ex_statics_2} and \ref{Ex_statics_3}, 2-dimensional random variables and their expectation vectors
 play important roles.

The set of positive integers and real numbers is denoted by $\NN$ and $\RR$, respectively.
For a $2$-dimensional vector $(x,y)\in\RR^2$, its Euclidean length is defined by $\Vert (x,y) \Vert:=\sqrt{x^2+y^2}$.
The coordinates of a vector $\bv$ in the $(x,y)$-plane are denoted by $v_x$ and $v_y$, respectively, i.e., $\bv=(v_x,v_y)$.
In the solutions, the number of significant digits is chosen to be 4 (for example, instead of $21.285$, we write $21.29$,
 or instead of $0.023456$, we write $0.02346$).
All of the physical quantities are expressed in SI units, so we do not display the units in calculations.
In the calculations, the magnitude of the gravitational acceleration near the Earth's surface is assumed to be 
 $g=9.810\,$$\mathrm{\frac{m}{s^2}}$.

\section{A stochastic approach in projectile motion}\label{section_projectule_motion}

In this section, we demonstrate our method for incorporating a stochastic point of view
 into physics exercises of mathematics education by presenting exercises for projectile motion.
In these ones, some of the initial values, namely, the initial height and/or velocity of a launched ball,
 are chosen randomly to make the problems more realistic.
We usually choose a uniform distribution, but in Exercise \ref{Ex_proj_motion_4}
 an absolutely continuous random variable with a given density function also comes into play
 for the ratio of the rebound and landing speeds when a tennis ball is dropped on concrete from a given distance.
In all of the exercises, we suppose that the air resistance is negligible.
Altogether, we present five problems in this section.
These exercises introduce a simple application of probability theory 
 to mechanical problems, they present composite systems influenced by random parameters. 
Students can see how variability in measurements can be modeled on a probabilistic basis.

\begin{Exe}\label{Ex_proj_motion_1}
Let us consider a cylindrical-shape forest lookout
 with a radius of 10 meters and a height of 50 meters, which stands on horizontal ground.
From the bottom of the forest lookout, Thomas goes up the spiral staircase.
Thomas stops at a height $h$ chosen according to a uniform distribution on the interval  $[0 \,\mathrm{m},50 \,\mathrm{m}]$, and
 from the edge of the forest lookout he launches a ball with velocity $v$ of which the magnitude is uniformly distributed on the interval
 $[5\,\mathrm{\frac{m}{s}}, 15\,\mathrm{\frac{m}{s}}]$, and its direction is perpendicular to the wall of the forest lookout.
We suppose that $h$ and $v$ are independent.
\begin{itemize}
\item[(i)] What is the expected value, variance and relative standard deviation of the distance of which the ball hits the ground from the wall of the forest lookout?

\item[(ii)] Suppose that the forest lookout is surrounded by an annulus-shape fence to which the top is horizontal.
     The two walls of the fence (inner and outer circles) are $10$ meters and $10.5$ meters from the wall of the forest lookout, respectively.
     The height of the fence is $2$ meters.
     On Figure \ref{Fig1_exe}, the planar section determined by the direction of the shot is shown,
     and only the part of the fence in the direction of the shot is plotted.
     What is the probability that the ball first hits one of the line segments $BC$ and $CD$ (see Figure \ref{Fig1_exe}),
     that is, the ball first hits the fence?
     \begin{figure}[ht]
       \centering
       \includegraphics[width=9cm]{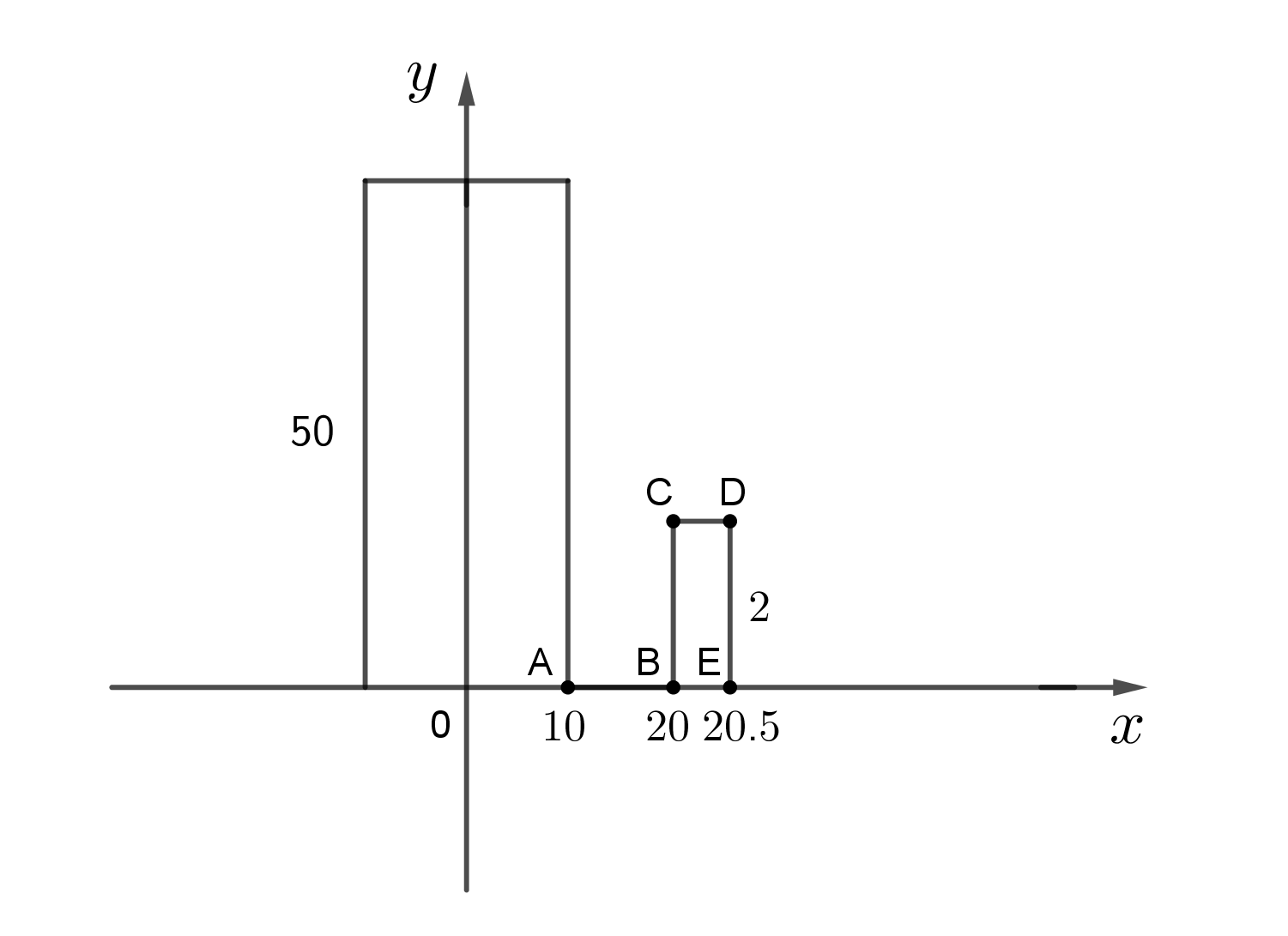}
       \caption{a planar section of the forest lookout surrounded by an annulus-shape fence for parts (ii) and (iii) in Exercise \ref{Ex_proj_motion_1}.}
       \label{Fig1_exe}
       \end{figure}

\item[(iii)]
Suppose that the forest lookout is surrounded by the same annulus-shape fence described in part (ii).
Further, suppose that the ball first hits the line segment $AB$ (that is,
 the ball first hits the ground in front of the fence but misses it).
What is the probability that Thomas launched the ball from lower than 10 meters?
\end{itemize}
\end{Exe}

\noindent{\bf Solution.}
It is known that if we launch a ball horizontally at height $h$ with velocity $v$, then
 the total (horizontal) distance of the whole journey before the ball hits the ground is
 $d:=\sqrt{\frac{2h}{g}}\cdot v$.
Further, if the ball takes a distance $x$ horizontally, then it takes a vertical distance
 $\frac{g}{2v^2}\cdot x^2$, see Figure \ref{Fig1} as well:
 \begin{figure}[ht]
 \centering
 \includegraphics[height=9cm]{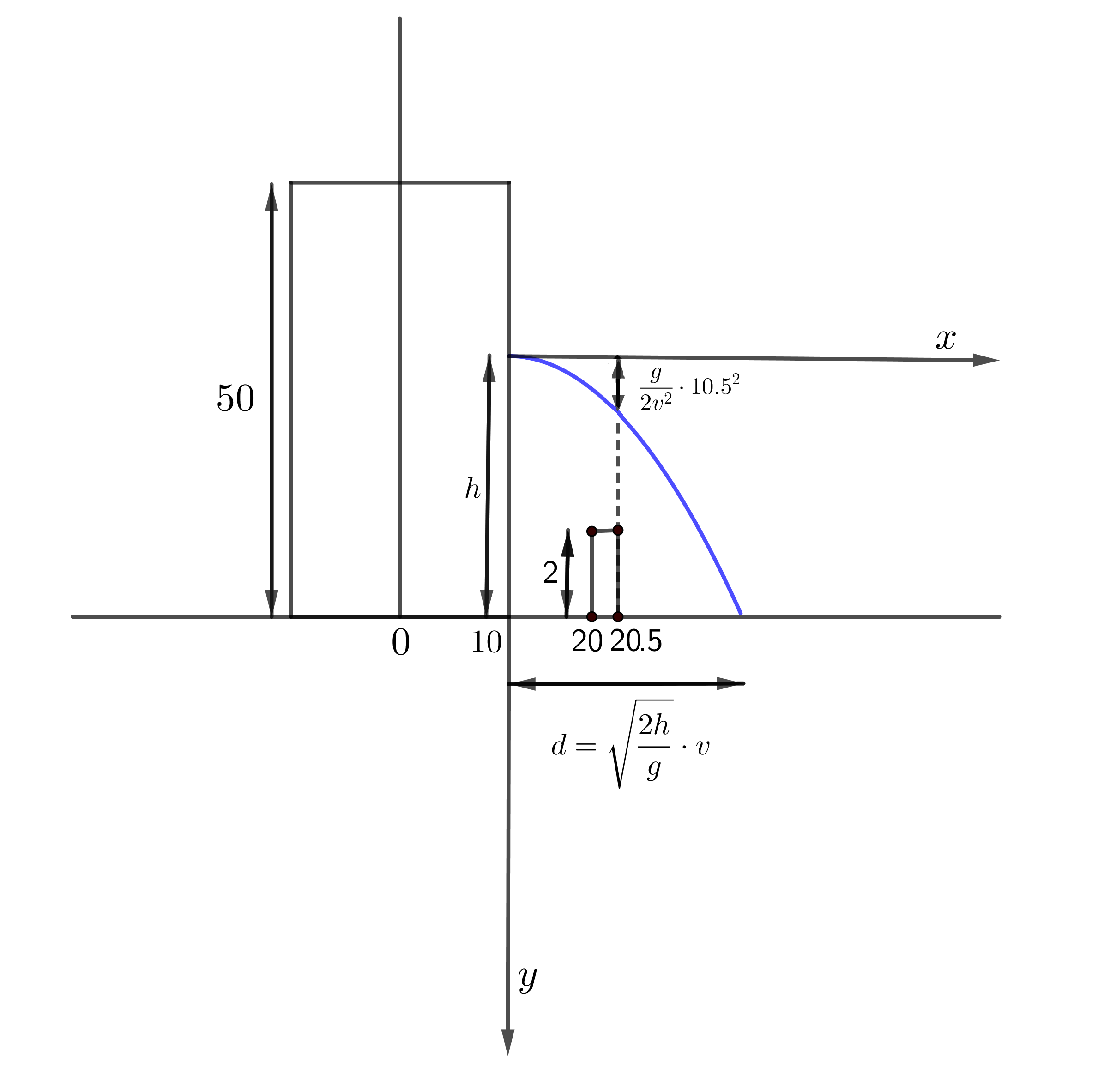}
  \caption{the trajectory of the motion of a ball launched horizontally at height $h$
             with velocity $v$ for the solution of Exercise \ref{Ex_proj_motion_1}.}
 \label{Fig1}
\end{figure}

By our assumptions, $h$ is a uniformly distributed random variable on the interval $[0,50]$,
 $v$ is a uniformly distributed random variable on the interval $[5,15]$,
 and furthermore, $h$ and $v$ are independent.

(i).
Then $\sqrt{h}$ and $v$ are also independent, hence, using the homogeneity of expectation (see \eqref{exp_linearity}),
 we get that
 \begin{align*}
   \EE(d)& = \EE\left(\sqrt{\frac{2h}{g}}\cdot v\right) = \sqrt{\frac{2}{g}}\,\EE(\sqrt{h})\cdot\EE(v)
          = \sqrt{\frac{2}{g}} \int_0^{50} \sqrt{p} \cdot \frac{1}{50}\,\dd p \cdot \int_5^{15} r \cdot\frac{1}{10}\,\dd r \\
         & =  \sqrt{\frac{2}{g}} \cdot \frac{1}{50}\cdot \frac{2\cdot 50^{3/2}}{3}\cdot \frac{1}{10}\cdot \frac{15^2-5^2}{2}
          = \frac{200}{3\sqrt{g}} \approx 21.29.
 \end{align*}
Now, we turn to calculate the variance and standard deviation of $d$.
Using that $h$ and $v^2$ are independent, we obtain that
 \begin{align*}
  \EE(d^2) & = \EE\left(\left( \sqrt{\frac{2h}{g}}\cdot v \right)^2\right)
             = \EE\left( \frac{2h}{g} \cdot v^2\right)
             = \frac{2}{g} \EE(h)\cdot\EE(v^2)
             = \frac{2}{g} \int_0^{50} p \cdot \frac{1}{50}\,\dd p \cdot \int_5^{15} r^2 \cdot\frac{1}{10}\,\dd r \\
           & = \frac{2}{g} \cdot \frac{1}{50} \cdot \frac{2500}{2}\cdot \frac{1}{10}\cdot \frac{3375 - 125}{3}
             = \frac{16250}{3g}.
 \end{align*}
Consequently, the variance of $d$ takes the form
 \[
   \DD^2(d) = \EE(d^2) - (\EE(d))^2 = \frac{16250}{3g} - \frac{40000}{9g}
            = \frac{8750}{9g}\approx 99.11,
 \]
 and hence the standard deviation of $d$ is
 \[
  \DD(d) = \frac{25\sqrt{14}}{3\sqrt{g}}\approx 9.955.
 \]
This implies that the relative standard deviation of $d$ is
 \[
   \rsd(d)=\frac{\frac{25\sqrt{14}}{3\sqrt{g}}}{\frac{200}{3\sqrt{g}}}
    = \frac{\sqrt{14}}{8} \approx 0.4677.
 \]

(ii).
Let us introduce the events $R$ and $S$:
 \begin{align*}
   &R:=\Big\{  \text{the ball's (first) point of impact is in the line segment $AB$} \Big\}
       = \Big\{ d<10 \Big\},\\
   &S:=\Big\{  \text{the ball's (first) point of impact is to the right of the point $E$} \Big\}\\
   &\phantom{:=\,}
      = \Big\{ \frac{g}{2v^2} \cdot 10.5^2 < h-2 \Big\}.
 \end{align*}
Then we get that
 \begin{align}\label{help1}
    \PP\big(\text{\{the ball hits the fence before it hits the ground\}}\big) = 1- \PP(R) - \PP(S).
 \end{align}

In what follows, we determine the probabilities $\PP(R)$ and $\PP(S)$ using the method of geometric probabilities.
In Appendix \ref{appAltSol}, as an alternative solution,  we determine these two probabilities using the law of total expectation as well.

\begin{figure}[ht]
 \centering
 \includegraphics[height=7cm]{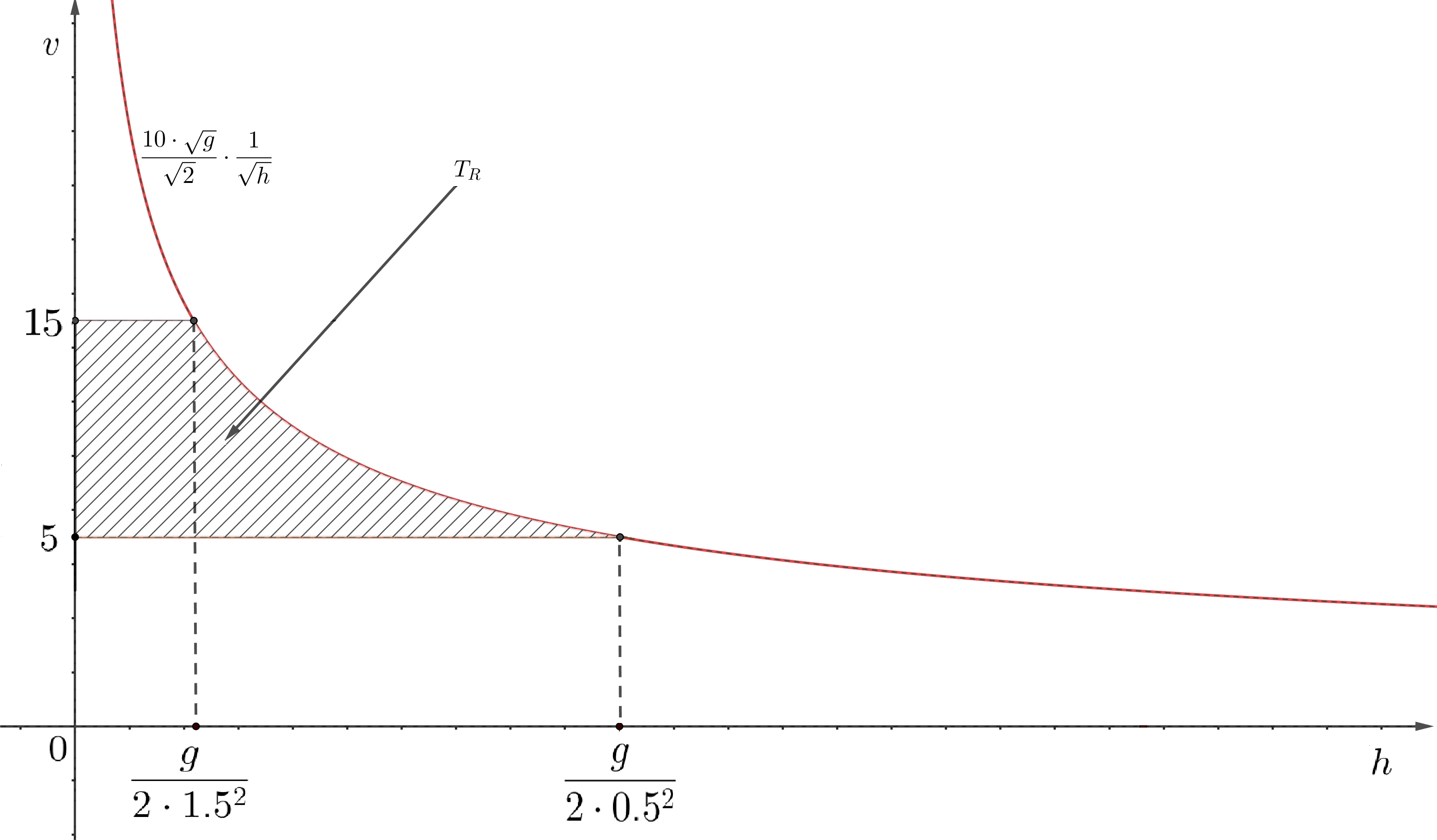}
 \caption{for calculating the probability $\PP(R)$  in the solution of part (ii) in Exercise \ref{Ex_proj_motion_1}.}
 \label{Fig1_a}
\end{figure}

By Figure \ref{Fig1_a}
 and the method of geometric probabilities we obtain that
 \begin{align*}
 \PP(R) &= \PP\left(d<10\right)
          = \PP\left(\sqrt{\frac{2h}{g}}\cdot v < 10\right)
                 = \PP\left( v < \frac{10\sqrt{g}}{\sqrt{2}}\cdot\frac{1}{\sqrt{h}}\right) \\
                &=\frac{T_R}{50(15-5)} = \frac{T_R}{500},
 \end{align*}
 where $500$ is the area of the rectangle $[0,50]\times [5,15]$ (''total'' area),
 and the ''probable'' area $T_R$ can be calculated as follows.
First, we determine the points at which the function $(0,\infty)\ni h\mapsto \frac{10\sqrt{g}}{\sqrt{2}}\cdot \frac{1}{\sqrt{h}}$
 takes the values $15$ and $5$, respectively (see Figure \ref{Fig1_a}).
By algebraic calculations, we can get that
 \begin{align}\label{help5}
  &\text{the solution of the equation \ $\frac{10\sqrt{g}}{\sqrt{2}}\cdot\frac{1}{\sqrt{h}} =15$, $h>0$, \ is \ $h=\frac{g}{2\cdot 1.5^2}  = 2.18$,}
 \end{align}
 and
 \begin{align}\label{help6}
  &\text{the solution of the equation \ $\frac{10\sqrt{g}}{\sqrt{2}}\cdot\frac{1}{\sqrt{h}} =5$, $h>0$, \ is \ $h=\frac{g}{2 \cdot 0.5^2} = 19.62$.}
 \end{align}
Consequently, by splitting the area $T_R$ into two parts, we have that
 \begin{align*}
  T_R&= \int_0^{\frac{g}{2\cdot 1.5^2}} (15-5)\,\dd p
       + \int_{\frac{g}{2 \cdot 1.5^2}}^{\frac{g}{2\cdot 0.5^2}} \left( \frac{10\sqrt{g}}{\sqrt{2}}\cdot\frac{1}{\sqrt{p}} - 5\right)\,\dd p \\
    & = \frac{10g}{2\cdot 1.5^2} + 10g\left(\frac{1}{0.5} - \frac{1}{1.5}\right)
       - \frac{5g}{2}\left(\frac{1}{0.5^2} - \frac{1}{1.5^2}\right)
      =\frac{20g}{3}.
 \end{align*}
Then
 \[
  \PP(R) = \frac{\frac{20g}{3}}{500} = \frac{g}{75} { = 0.1308.}
 \]

\begin{figure}[ht]
 \centering
 \includegraphics[height=7cm]{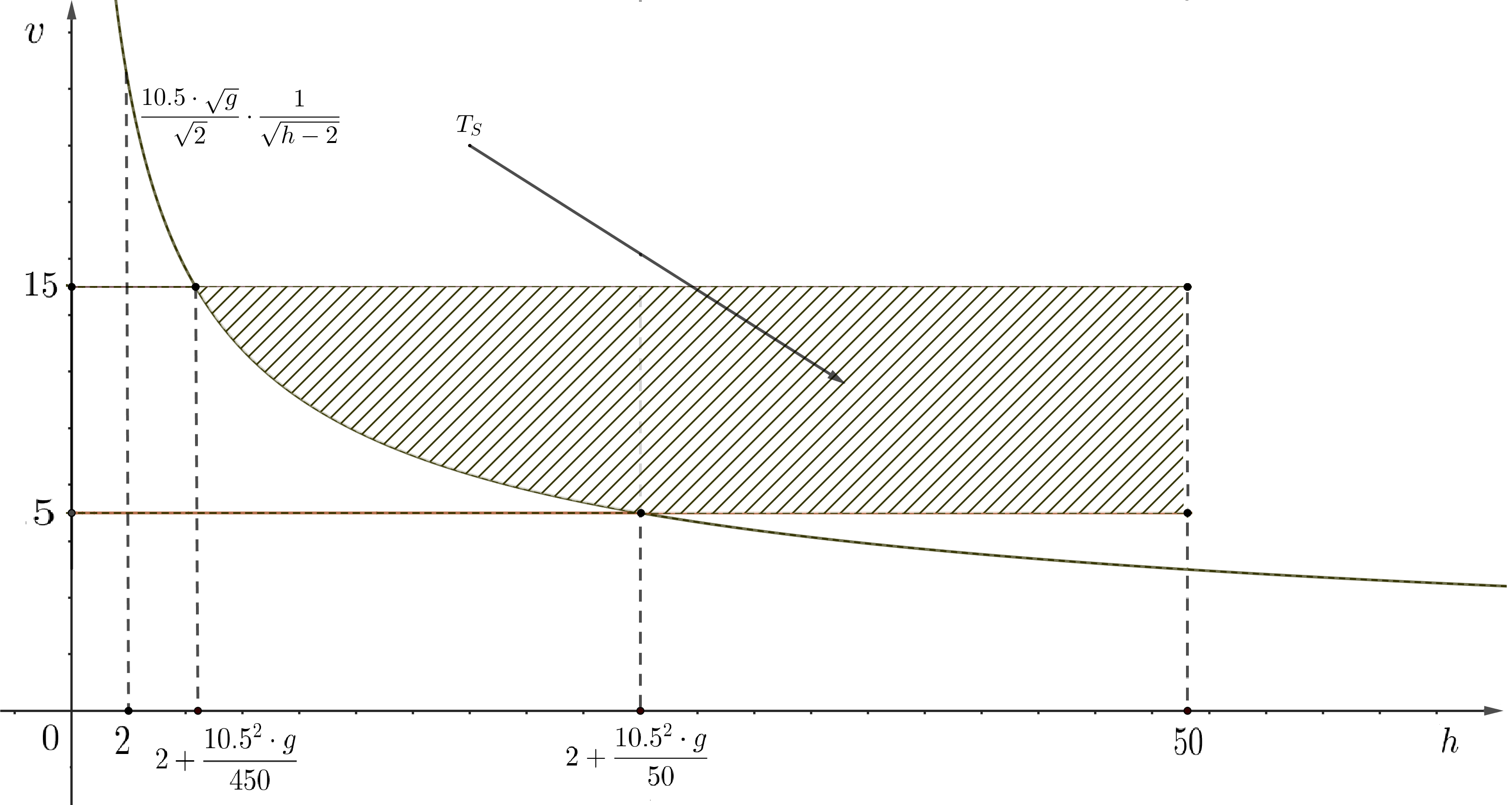}
 \caption{for calculating the probability $\PP(S)$  in the solution of part (ii) in Exercise \ref{Ex_proj_motion_1}.}
 \label{Fig1_b}
\end{figure}

Similarly, we obtain
 \begin{align*}
 \PP(S)& = \PP\left(\frac{g}{2v^2}\cdot 10.5^2 < h-2\right)
        =  \PP\left(\frac{g\cdot 10.5^2}{2(h-2)} < v^2, h > 2\right) \\
       & =  \PP\left( v > \frac{10.5 \sqrt{g}}{\sqrt{2}}\cdot \frac{1}{\sqrt{h-2}}, \, h > 2\right)
        = \frac{T_S}{500},
 \end{align*}
 where the ''probable'' area $T_S$ can be caclulated as follows.
First, we determine the points at which the function
  $(2,\infty) \ni h \mapsto \frac{10.5\sqrt{g}}{\sqrt{2}}\cdot \frac{1}{\sqrt{h-2}}$
 takes the values $15$ and $5$, respectively (see Figure \ref{Fig1_b}).

By algebraic calculations, we can get that
 \begin{align*}
  &\text{the solution of the equation \ $\frac{10.5\sqrt{g}}{\sqrt{2}}\cdot\frac{1}{\sqrt{h-2}} =15$, $h>2$, \ is \ $h=2+\frac{10.5^2g}{450} \approx 4.404$,}
 \end{align*}
 and
 \begin{align*}
  &\text{the solution of the equation \ $\frac{10.5\sqrt{g}}{\sqrt{2}}\cdot\frac{1}{\sqrt{h-2}} =5$, $h>2$, \ is \ $h=2+\frac{10.5^2g}{50} \approx  23.63$.}
 \end{align*}
Then
  \begin{align*}
 T_S&= \int_{2+\frac{10.5^2 g}{450}}^{2+\frac{10.5^2g}{50}} \left( 15 - \frac{10.5\sqrt{g}}{\sqrt{2}}\cdot\frac{1}{\sqrt{p-2}} \right)\,\dd p
       + \int_{2+\frac{10.5^2g}{50}}^{50} (15-5)\,\dd p \\
       & = 15 \left( \frac{10.5^2g}{50} - \frac{10.5^2g}{450} \right)
        - 10.5\sqrt{2g}\left( \frac{10.5 \sqrt{g}}{\sqrt{50}} - \frac{10.5\sqrt{g}}{\sqrt{450}} \right)
         +10\left(50-2 - \frac{10.5^2g}{50} \right) \\
       & = 480 - \frac{10.5^2g}{15}.
 \end{align*}
Consequently,
 \[
  \PP(S) = \frac{480 - \frac{10.5^2g}{15}}{500}
         = \frac{48}{50} - \frac{110.25 g}{7500}\approx 0.8158.
 \]

Using the values of $\PP(R)$ and $\PP(S)$, \eqref{help1} implies that
 \begin{align*}
   1- \PP(R) - \PP(S) = 1 -  \frac{g}{75}  - \frac{48}{50} + \frac{110.25 g}{7500}
                      = \frac{2}{50} + \left(\frac{110.25}{7500} - \frac{1}{75}\right)g \approx 0.05341.
 \end{align*}

\begin{figure}[ht]
 \centering
 \includegraphics[height=7cm]{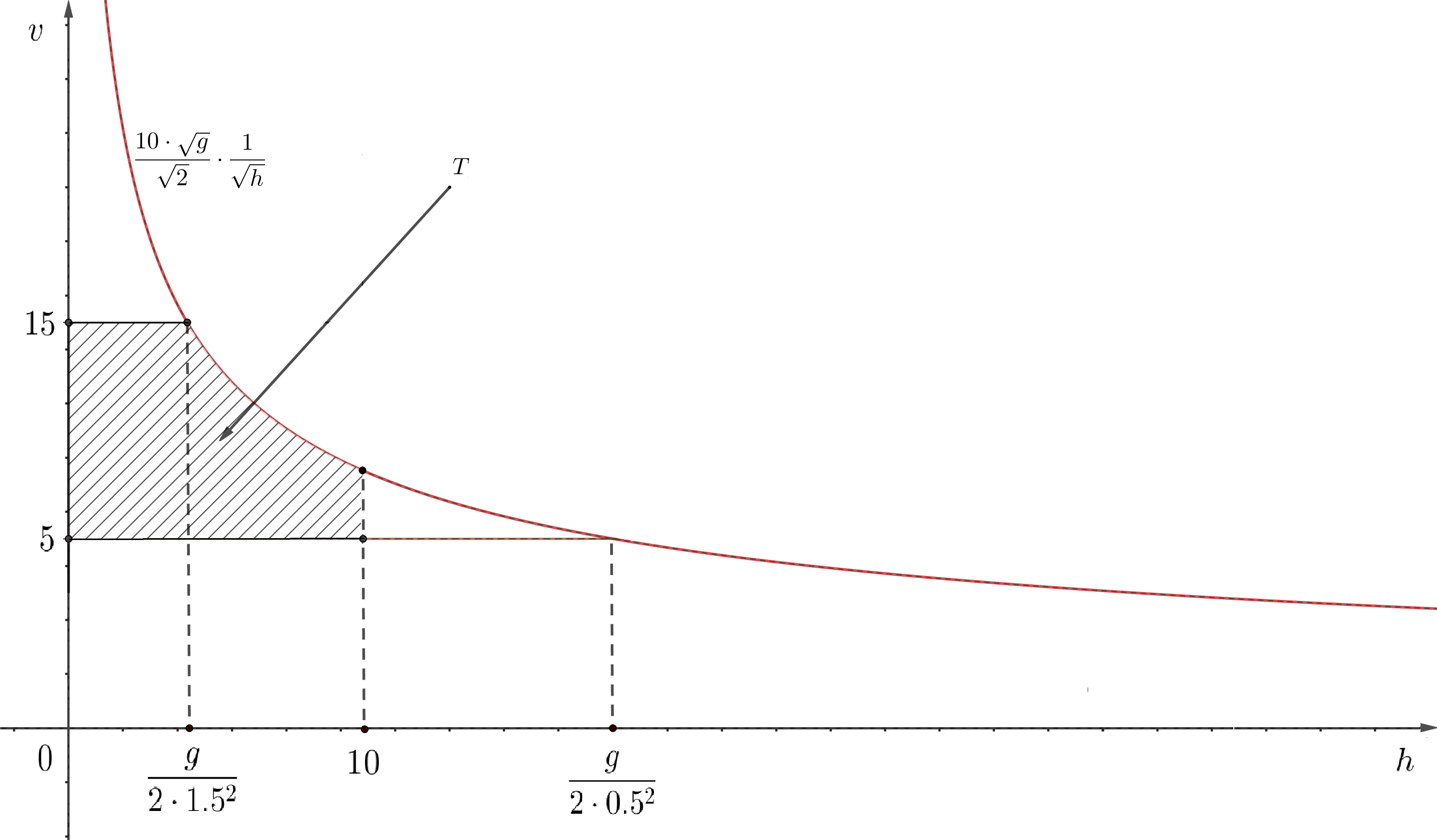}
 \caption{for calculating the probability $\PP(h<10, d< 10)$ in the solution of part (iii) in Exercise \ref{Ex_proj_motion_1}.}
 \label{Fig2}
\end{figure}

(iii).
We have to calculate the conditional probability $\PP(h<10 \mid d<10)$.
By part (ii), we have that $\PP(d<10)=\frac{9}{75}$, and hence, by the definition of conditional probability,
 \begin{align*}
   \PP(h<10 \mid d<10) = \frac{\PP(h<10,d<10)}{\PP(d<10)}
                       = \frac{75}{g} \PP(h<10,d<10).
 \end{align*}
Using Figure \ref{Fig2}
 and the method of geometric probabilities, we get that
 \begin{align*}
 \PP(h<10,d<10) &= \PP\left(h<10, \sqrt{\frac{2h}{g}}\cdot v < 10\right)
                 = \PP\left(h<10, v < \frac{10\sqrt{g}}{\sqrt{2}}\cdot\frac{1}{\sqrt{h}}\right) \\
                &=\frac{T}{50(15-5)} = \frac{T}{500},
 \end{align*}
 where $500$ is the area of the rectangle $[0,50]\times [5,15]$ (''total'' area),
 and the ''probable'' area $T$ can be calculated as follows.
Using \eqref{help5} and \eqref{help6}, we have that
  \begin{align*}
   T& =\int_0^{\frac{g}{2\cdot 1.5^2}} (15-5)\,\dd p
       + \int_{\frac{g}{2\cdot 1.5^2}}^{10} \left(  \frac{10\sqrt{g}}{\sqrt{2}}\cdot\frac{1}{\sqrt{p}} - 5\right)\,\dd p\\
    &=\frac{10g}{2\cdot 1.5^2} + 10\sqrt{2g}\left( \sqrt{10} - \frac{\sqrt{g}}{1.5\sqrt{2}}\right) - 5\left(10- \frac{g}{2\cdot 1.5^2}\right)\\
    &= \frac{10g}{4.5} + 10\sqrt{20g} - \frac{10g}{1.5} - 50 + \frac{5g}{4.5}\\
    &= -\frac{15g}{4.5} + 10\sqrt{20g}-50.
  \end{align*}

We remark that the probability $\PP(h<10,d<10)$ could be determined using the law of total expectation as well
 similarly as in the solution of part (ii) (by Method 2), but we will leave it for now.

All in all, we get that
 \begin{align*}
   \PP(h<10 \mid d<10) = \frac{75}{500}\left( -\frac{15}{4.5} + \frac{10\sqrt{20}}{\sqrt{g}} - \frac{50}{g} \right)
                       \approx 0.8772.
 \end{align*}
\proofend

\begin{Rem}\label{Rem_Exercise21}
 (i).\
We compare the solution of part (i) of Exercise \ref{Ex_proj_motion_1} with the case when
 the height (where Thomas stops) is chosen to be $25\,\mathrm{m}$ and the magnitude of the velocity of the ball
 is chosen to be $10\,\mathrm{\frac{m}{s}}$.
Note that the values $25$ and $10$ are nothing else but the expected values of the corresponding uniformly
 distributed random variables considered in Exercise \ref{Ex_proj_motion_1} for the height and
 magnitude of velocity, respectively.
Then the total (horizontal) distance of the whole journey before the ball hits the ground is
 $\sqrt{\frac{2\cdot 25}{g}}\cdot 10\approx 22.58$.
We call the attention to the fact that $\EE(d)$ calculated in the solution of part (i) in Exercise \ref{Ex_proj_motion_1}
 is not equal to  $\sqrt{\frac{50}{g}}\cdot 10$.
The relative error of the two quantities in question is
\[
  \frac{\EE(d)}{\sqrt{\frac{50}{g}}\cdot 10}
     = \frac{\frac{200}{3\sqrt{g}}}{\sqrt{\frac{50}{g}}\cdot 10}
     = \frac{2\sqrt{2}}{3}<1.
\]

(ii).\
We note that students may formulate and solve some natural counterparts of Exercise \ref{Ex_proj_motion_1}.
Namely, one could choose the height (where Thomas stops) according to a (discrete) uniform distribution, for example,
 on the finite set $\{1\,\mathrm{m},2\,\mathrm{m},\ldots, 49\,\mathrm{m}, 50\,\mathrm{m}\}$.
Of course, one could also consider other discrete or absolutely continuous distributions
with range in $[0,50]$. For example, the discrete distribution on the set $\{10\,\mathrm{m},20\,\mathrm{m},30\,\mathrm{m},40\,\mathrm{m},50\,\mathrm{m}\}$
with corresponding probabilities $\frac{40}{100}, \frac{30}{100}, \frac{20}{100},\frac{9}{100},\frac{1}{100}$
would may suggest that Thomas has a fear of heights.
As for another variant of Exercise \ref{Ex_proj_motion_1}, one could consider a case when only one of the inputs is randomized,
 for example, one could choose the height (where Thomas stops) according to some probability distribution, but the magnitude
 of the velocity of the ball is taken to be a fixed (non-random) value.
\proofend
\end{Rem}

\begin{Exe}\label{Ex_proj_motion_2}
Let us consider a rectangular cuboid-shape building with a height of 50 meters, standing on horizontal ground.
From the bottom of the building, Anne takes the elevator up.
Anne stops at a height $h$ chosen according to a uniform distribution on the interval  $[0 \,\mathrm{m},50 \,\mathrm{m}]$,
 and from the edge of the building she throws a ball with velocity $v$
 of which the magnitude is $15\,$$\mathrm{\frac{m}{s}}$
 and its direction is perpendicular to the wall of the building.
Parallel to Anne's building, there is another rectangular cuboid-shape building, 15 metres away, its height is $50$ meters.
If the thrown ball reaches the opposite house, then it bounces back elastically from its wall,
 and if the rebounced ball also reaches the wall of Anne's house, then it rebounces (elastically) from it as well, and so on.
 \begin{itemize}
  \item[(i)] What is the distribution of the number of times the ball rebounces before it hits the ground?
  \item[(ii)] What is the expectation, variance and relative standard deviation of the number of times the ball rebounces before it hits the ground?
 \item[(iii)] What is the probability that the ball hits the ground closer to Anne's building?
 \end{itemize}
\end{Exe}

\noindent{\bf Solution.}
(i).
It is known that if we launch a ball horizontally at height $h$ with velocity $v$,
 then the total (horizontal) distance of the whole journey before the ball hits the ground is $d:=\sqrt{\frac{2h}{g}}\cdot v$.

Since we assumed that the rebounds (collisions) are elastic, the trajectory after the first collision (provided that such a collision occurs)
 is the reflection of the imaginary continuation of the trajectory before the first collision after the colliding surface (wall) on the colliding surface.
Similarly, the trajectory after the second collision (provided that such a collision occurs) is the reflection of the
 imaginary continuation of the trajectory after the first collision after the second colliding surface on the (second) colliding surface.
This train of thought can be continued in a similar way for all of the collisions that occur.
Consequently, it is enough to consider a single projectile motion, launched at height $h$ with horizontal velocity $v$.

Let the random variable $\xi$ denote the number of times the ball rebounces before it hits the ground.

For all nonnegative integer $k$, we have that
 \begin{align*}
   &\{\xi = k\}
      = \Big\{ \text{ the distance $d$ traveled horizontally by the ball before it hits the ground}\\
   &\phantom{\{\xi = k\} = \Big\{\,}
            \text{ is in $[15k, 15(k+1))$ } \Big\}.
 \end{align*}
Since the maximum value of $d$ is $\sqrt{\frac{2\cdot50}{g}}\cdot 15\approx 47.89$, the ball can rebound maximum three times.
Hence the range of $\xi$ is $\{0,1,2,3\}$, and, using that $h$ is a uniformly distributed random variable
 on the interval $[0,50]$, we get that
 \begin{align*}
   &\PP(\xi=0) = \PP(d<15) = \PP\left(\sqrt{\frac{2h}{g}}\cdot 15 < 15\right) = \PP(h< g/2)=\frac{g/2}{50} = \frac{g}{100},\\
   &\PP(\xi=1) = \PP(15\leq d<30) = \PP\left(15\leq \sqrt{\frac{2h}{g}}\cdot 15 < 30\right)
               = \PP\left(\frac{g}{2}\leq h< \frac{4g}{2} \right)
              = \frac{3g}{100},\\
   &\PP(\xi=2) = \PP(30\leq d<45) = \PP\left(30\leq \sqrt{\frac{2h}{g}}\cdot 15 < 45\right)
               = \PP\left(\frac{4g}{2} \leq h< \frac{9g}{2}\right)
               =\frac{5g}{100},\\
   &\PP(\xi=3) = \PP(45\leq d<65) = 1- \PP(d<45) = 1 - (\PP(\xi=0) + \PP(\xi=1) + \PP(\xi=2))\\
   &\phantom{\PP(\xi=3)}
              = 1 - \frac{9g}{100}.
 \end{align*}

(ii).
The expected value of $\xi$ takes the form
 \begin{align*}
   \EE(\xi)& = \sum_{k=0}^3 k\PP(\xi=k)
            = 1\cdot \frac{3g}{100} +  2\cdot \frac{5g}{100} +  3\cdot \frac{100-9g}{100}
            = \frac{3g+10g+300-27g}{100}\\
           & = \frac{300-14g}{100}  \approx 1.627.
 \end{align*}
Now, we turn to calculate the variance and standard deviation of $\xi$.
We have that
 \begin{align*}
   \EE(\xi^2)& = \sum_{k=0}^3 k^2\PP(\xi=k)
            = 1\cdot \frac{3g}{100} +  4\cdot \frac{5g}{100} +  9\cdot \frac{100-9g}{100}
            = \frac{3g+20g+900-81g}{100}\\
           & = \frac{900-58g}{100}  \approx 3.3102.
 \end{align*}
Consequently, the variance of $\xi$ takes the form
 \begin{align*}
    \DD^2(\xi) = \EE(\xi^2) - (\EE(\xi))^2
               = \frac{900-58g}{100} - \left(\frac{300-14g}{100}\right)^2
               = \frac{2600g - 196g}{10000}
               \approx 0.6644,
 \end{align*}
 and hence the standard deviation of $\xi$ is
 \[
 \DD(\xi) = \sqrt{ \frac{2600g - 196g^2}{10000} }
            = \frac{\sqrt{650g-49g^2}}{50}
            \approx 0.8151.
 \]
This implies that the relative standard deviation of $\xi$ is
 \[
   \rsd(\xi)=\frac{ \frac{\sqrt{650g-49g^2}}{50} }{ \frac{300-14g}{100} }
      = \frac{\sqrt{650g-49g^2}}{150-7g}\approx 0.5011.
 \]

(iii).
Let $A$ be the event that the ball hits the ground closer to Anne's building.
Using the additivity of probability and that the maximum value of $d$ is
 $\sqrt{\frac{2\cdot50}{g}}\cdot 15\approx 47.89$ (see part (i)), we get that
 \begin{align*}
   \PP(A) &= \sum_{k=0}^3 \PP(A,\xi=k)\\
          &= \PP(d<7.5) + \PP(15+7.5 \leq d< 15+15) + \PP(30 \leq d\leq 30+7.5) \\
          &\phantom{=\;} + \PP(45+7.5 \leq  d< 45+15) \\
          &= \PP(d<7.5) + \PP(22.5 \leq d< 37.5) + 0\\
          &= \PP\left(  \sqrt{\frac{2h}{g}}\cdot 15 < 7.5 \right)
             + \PP\left(  22.5 \leq \sqrt{\frac{2h}{g}}\cdot 15 < 37.5 \right) \\
          &= \PP\left(h \leq \frac{g}{8}\right) + \PP\left( \frac{9g}{8} \leq h < \frac{25g}{8} \right)
           = \frac{1}{50}\left( \frac{g}{8} + \frac{(25-9)g}{8}\right)
           =\frac{17g}{400} \approx 0.4169.
 \end{align*}
\proofend

Part (i) of the next Exercise \ref{Ex_proj_motion_3} is a randomized version of Exercise 2 for 4-th grade high school students
 at the \'Agoston Bud\'o Physics Competition (1980) (see \cite{Bud}) in the sense that we randomly choose the height from which we drop the ball.
Parts (ii), (iii) and (iv) of Exercise \ref{Ex_proj_motion_3} contain new probabilistic-type questions
 to the same phenomena.

\begin{Exe}\label{Ex_proj_motion_3}
From a height chosen uniformly distributed between $0$ and $1$ meters, we drop a ball on the table.
We assume that the ball loses $5\%$ of its velocity by each collision with the table.
\begin{itemize}
 \item[(i)] What is the density function, expectation, variance and relative standard deviation of the total time (in seconds) that elapses before the ball stops moving?

 \item[(ii)] What is the density function, expectation, variance and relative standard deviation of the total distance (in meters) that the ball travels during its motion?

 \item[(iii)] What is the probability that the ball ends its movement earlier than an identical ball dropped
              on the table from a height of $0.5$ meters?
              Calculate the same probability if $0.5$ meters is replaced by $\frac{4}{9}$ meters!
              What could be the reason for choosing $\frac{4}{9}$ meters in the previous task?

 \item[(iv)] Let $\xi$ denote a random variable that counts the number of times the ball is at a height of $0.1$ meter during its movement.
             Let us determine the distribution and expectation of $\xi$!
\end{itemize}
\end{Exe}

\noindent{\bf Solution.}
Let $h_0$ denote the number of metres from which the ball is dropped.
Then $h_0$ is uniformly distributed on the interval $[0,1]$.
Let us introduce the following notations: for each nonnegative integer $n$, let
 $v_{n+1}:=0.95\,v_n$, where $v_0$ is the velocity of a ball dropped
 from a height of $h_0$ when it hits the ground.
The movement that takes place (bouncing) is an infinite sequence of consecutive free falls and vertical upward projectile motions.
A ball dropped from a height of $h_0$ takes time $t_0:=\sqrt{\frac{2h_0}{g}}$ to hit the ground, and then its velocity is
 $v_0=gt_0=\sqrt{2gh_0}$.

By the assumption, the initial velocity of the vertical upward projectile motion after the first bounce is $v_1=0.95\, v_0$.
It is known that the time to reach the maximum height is $t_1:=\frac{v_1}{g}$ and the maximum height of the projectile motion
 in question is $\frac{v_1^2}{2g}$.
The new (second) fall time is $t_1':=t_1$, since there is no energy loss, and the velocity
 at the (second) landing is $\sqrt{2g\frac{v_1^2}{2g}}=v_1$
 (that is, it is equal to the initial velocity of the vertical upward projectile motion after the first rebounce).

Similarly, for the vertical upward projectile motion after the second rebounce, by the assumption, we have that
 its initial velocity is $v_2=0.95 v_1=0.95^2 v_0$,
 the time to reach its maximum height is $t_1:=\frac{v_1}{g}$,
 its maximum height is $\frac{v_2^2}{2g}$,
 and its fall time is $t_2':=t_2$, since there is no energy loss.

By induction, for the vertical upward projectile motion after the $n^\mathrm{th}$ rebounce, by the assumption, we have that
 its initial velocity is $v_n=0.95^n v_0$,
 the time to reach its maximum height is $t_n:=\frac{v_n}{g} = \frac{0.95^nv_0}{g}$,
 its maximum height is $\frac{v_n^2}{2g}$,
 and its fall time is $t_n':=t_n$, since there is no energy loss.

Consequently, the total time that elapses before the ball stops moving is
 \begin{align*}
   t:=t_0+t_1+t_1'+t_2+t_2'+ \cdots=t_0+\sum_{n=1}^\infty (t_n+t_n') = t_0+2\sum_{n=1}^\infty t_n,
 \end{align*}
 and the total distance that the ball travels during its motion is
 \[
  s:=h_0 +2\frac{v_1^2}{2g} + 2\frac{v_2^2}{2g} + \cdots = h_0 + \frac{1}{g}\sum_{n=1}^\infty v_n^2.
 \]
Using these two formulas, we get that
 \begin{align*}
  t&=\sqrt{\frac{2h_0}{g}} + 2\sum_{n=1}^\infty \frac{0.95^n v_0}{g}
    = \sqrt{\frac{2h_0}{g}} + \frac{2}{g} \sqrt{2g h_0}\sum_{n=1}^\infty 0.95^n
    = \sqrt{\frac{2h_0}{g}} + 2\sqrt{\frac{2h_0}{g}} \frac{0.95}{1-0.95}\\
   &=  39 \sqrt{\frac{2}{g}}\sqrt{h_0},
 \end{align*}
 and
 \begin{align*}
  s=h_0+\frac{1}{g}\sum_{n=1}^\infty (0.95^n v_0)^2
   = h_0 + \frac{v_0^2}{g}\sum_{n=1}^\infty 0.95^{2n}
   = h_0 + 2h_0\frac{0.95^2}{1-0.95^2}
  = \frac{761}{39} h_0.
 \end{align*}

(i).
First, we determine the cumulative distribution function of $t$.
If $x\leq 0$, then $\PP(t<x)=0$; if $x\geq 39\sqrt{\frac{2}{g}}$, then $\PP(t<x)=1$; and
 if $x\in\Big(0,39\sqrt{\frac{2}{g}}\Big)$, then we have that
 \begin{align*}
   \PP(t<x) = \PP\left(39\sqrt{\frac{2}{g}}\sqrt{h_0} < x\right)
            = \PP\left( h_0 < \frac{g}{3042} x^2\right)
            = \frac{g}{3042} x^2.
 \end{align*}
Consequently, the cumulative distribution function $F_t:\RR\to[0,1]$ of $t$ takes the form
 \begin{align*}
  F_t(x) = \begin{cases}
              0 & \text{if $x\leq 0$,}\\[1mm]
              \frac{g}{3042} x^2 & \text{if $x\in\Big(0,39\sqrt{\frac{2}{g}}\Big)$,}\\[1mm]
              1 & \text{if $x\geq 39\sqrt{\frac{2}{g}}$.}
          \end{cases}
 \end{align*}
This implies that the density function $f_t:\RR\to[0,\infty)$ of $t$ takes the form
 \begin{align*}
  f_t(x) = \begin{cases}
              \frac{g}{1521} x & \text{if $x\in\Big(0,39\sqrt{\frac{2}{g}}\Big)$,}\\
              0 & \text{otherwise.}
          \end{cases}
 \end{align*}

The expected value of the total time $t$ that elapses before the ball stops moving is
 \begin{align*}
    \EE(t) = 39 \sqrt{\frac{2}{g}} \EE(\sqrt{h_0})
           = 39 \sqrt{\frac{2}{g}} \int_0^1 \sqrt{x}\,\dd x
           = 39 \sqrt{\frac{2}{g}}\cdot \frac{2}{3}
           \approx 11.74.
 \end{align*}
Alternatively, we can also calculate the expected value of $t$ using its density function $f_t$ as follows:
 \begin{align*}
   \EE(t) = \int_0^{39\sqrt{\frac{2}{g}}} x\cdot \frac{g}{1521} x\,\dd x
         = \frac{g}{1521}\cdot \frac{\left(39\sqrt{\frac{2}{g}}\right)^3}{3}
         = 39 \sqrt{\frac{2}{g}}\cdot \frac{2}{3}.
 \end{align*}

We call the attention to the fact that $\EE(t)$ is not equal to
  the total time that elapses before a ball dropped from a height of $0.5$ meters stops moving
 $\Big(39 \sqrt{\frac{2}{g}}\sqrt{0.5}\approx 12.45\Big)$,
 but it is equal to the total time that elapses before a ball dropped
 from a height of $\frac{4}{9}$ stops moving.

Now, we turn to calculate the variance and standard deviation of $t$.
We have that
  \begin{align*}
    \EE(t^2) = 39^2 \cdot\frac{2}{g} \EE(h_0)
             = 39^2 \cdot\frac{2}{g} \cdot \frac{1}{2}
             = \frac{39^2}{g},
  \end{align*}
 and hence the variance and standard deviation of $t$ take the following forms
 \[
  \DD^2(t) = \frac{39^2}{g} - 39^2\cdot \frac{2}{g}\cdot \frac{4}{9}
           = \frac{39^2}{9g} \approx 17.23,
 \]
 and
 \[
  \DD(t)= \sqrt{\frac{39^2}{9g}}.
 \]
This implies that the relative standard deviation of $t$ is
 \[
    \rsd(t)=\frac{\frac{39}{3\sqrt{g}}}{39 \sqrt{\frac{2}{g}}\cdot \frac{2}{3}}
             = \frac{1}{2\sqrt{2}}\approx 0.3536.
 \]

(ii).
Since $s=\frac{761}{39} h_0$ and $h_0$ is uniformly distributed on the interval $[0,1]$, we get that
 $s$ is uniformly distributed on the interval $\big[0,\frac{761}{39}\big]$, and its density function $f_s:\RR\to [0,\infty)$
 takes the form
 \begin{align*}
  f_s(x) = \begin{cases}
              \frac{39}{761} & \text{if $x\in\big[0,\frac{761}{39}\big]$,}\\
              0 & \text{otherwise.}
          \end{cases}
 \end{align*}

The expected value of the total distance $s$ that the ball travels during its motion is
 \[
  \EE(s) = \frac{761}{39} \EE(h_0)
         = \frac{761}{39} \cdot \frac{1}{2}
         = \frac{761}{78}\approx  9.756.
 \]
Note that $\EE(s)$ is equal to the total distance that a ball dropped from a height of $0.5$ meters
 travels during its motion.

Now, we turn to calculate the variance and standard deviation of $s$.
We have that
  \begin{align*}
    \EE(s^2) = \int_0^{761/39} x^2 \cdot \frac{39}{761}\,\dd x
             = \frac{39}{761}\cdot \frac{1}{3} \left( \frac{761}{39}\right)^3
             = \frac{1}{3} \left( \frac{761}{39}\right)^2,
  \end{align*}
 and hence the variance and standard deviation of $s$ take the following forms
 \[
  \DD^2(s) =  \frac{1}{3} \left( \frac{761}{39}\right)^2  - \left( \frac{761}{78}\right)^2
           =  \frac{1}{12} \left( \frac{761}{39}\right)^2 \approx 31.73,
 \]
 and
 \[
  \DD(s)= \sqrt{ \frac{1}{12} \left( \frac{761}{39}\right)^2} = \frac{761}{78\sqrt{3}}.
 \]
This implies that the relative standard deviation of $s$ is
 \[
    \rsd(s)=\frac{\frac{761}{78\sqrt{3}}}{ \frac{761}{78} }
             = \frac{1}{\sqrt{3}}\approx 0.5774.
 \]

Since the relative standard deviation of $s$ is greater than that of $t$, roughly speaking, one can say that
 the level of dispersion of the values of $s$ around it mean $\EE(s)$ is greater than that of the values of $t$ around its mean $\EE(t)$.

(iii). Then we have that
 \[
    \PP\Big( 39 \sqrt{\frac{2}{g}}\sqrt{h_0} < 39 \sqrt{\frac{2}{g}}\sqrt{0.5}\Big)
       = \PP(\sqrt{h_0}< \sqrt{0.5}) = \PP(h_0<0.5) = \frac{1}{2},
 \]
 and
 \[
    \PP\left( 39 \sqrt{\frac{2}{g}}\sqrt{h_0} < 39 \sqrt{\frac{2}{g}}\sqrt{\frac{4}{9}}\right)
       = \PP\left(\sqrt{h_0}< \sqrt{\frac{4}{9}}\right) = \PP\left(h_0<\frac{4}{9}\right) = \frac{4}{9}\approx 0.4444.
 \]

The reason for choosing $\frac{4}{9}$ is that, in the solution of parts (i) and (ii), we have seen that
 the expectation of the total time that elapses before the ball stops moving is equal to
 the total time that elapses before a ball dropped from a height of $\frac{4}{9}$ stops moving, that is,
 $\EE\left( 39 \sqrt{\frac{2}{g}} \sqrt{h_0}\right) = 39 \sqrt{\frac{2}{g}} \sqrt{\frac{4}{9}}$.
This also provides an example for the fact that given two random variables $\eta$ and
 $\zeta$, it can happen that their expected values are equal, i.e., $\EE(\eta)=\EE(\zeta)$, but $\PP(\eta<\zeta)\ne \frac{1}{2}$.

(iv).
The range of the random variable $\xi$ is $\{0,1,2,\ldots\}$, but we will check that
 $\xi$ takes every positive even number with probability $0$, and it will also turn out that
 there exists a natural number $n_0$ such that the probability that $\xi$ takes
 a value greater than $n_0$ is $0$.

Using the calculations before parts (i) and (ii), we have that $\PP(\xi=0)=\PP(h_0<0.1)=0.1$, furthermore,
 \begin{align*}
   \PP(\xi=1) &= \PP\Big(h_0\geq 0.1\,, \frac{v_1^2}{2g} < 0.1\Big)
              = \PP\Big(h_0\geq 0.1\,, \frac{(0.95 v_0)^2}{2g} < 0.1\Big) \\
              &= \PP\Big(h_0\geq 0.1\,, \frac{0.95^2 \cdot 2gh_0}{2g} < 0.1\Big)
              = \PP\Big(0.1\leq h_0 < \frac{0.1}{0.95^2} \Big)\\
              &=\frac{0.1}{0.95^2} - 0.1
              \approx 0.01080,
 \end{align*}
 and, since $h_0$ is has an absolutely continuous distribution,
 \[
    \PP(\xi=2) = \PP\Big(h_0\geq 0.1\,, \frac{v_1^2}{2g} = 0.1\Big) = \PP\Big( h_0 > 0.1\,, h_0= \frac{0.1}{0.95^2} \Big) = 0.
 \]
Similarly,
 \begin{align*}
   \PP(\xi=3) & = \PP\Big(h_0 > 0.1\,, \frac{v_1^2}{2g} > 0.1\,, \frac{v_2^2}{2g} < 0.1\Big)
                = \PP\Big(h_0 > 0.1\,,  h_0 > \frac{0.1}{0.95^2}, \frac{(0.95^2 v_0)^2}{2g} < 0.1  \Big)\\
              & = \PP\Big(h_0 > 0.1\,,  h_0 > \frac{0.1}{0.95^2}, \frac{0.95^4\cdot 2g h_0}{2g} < 0.1  \Big)\\
              & = \PP\Big(\frac{0.1}{0.95^2} < h_0 < \frac{0.1}{0.95^4} \Big)
                =\frac{0.1}{0.95^2}\cdot \frac{1-0.95^2}{0.95^2}.
 \end{align*}
One can also see that
 \[
   \PP(\xi=3) = \frac{\PP(\xi=1)}{0.95^2} > \PP(\xi=1).
 \]

In what follows, let $k\in\{1,2,\ldots\}$.
Since $h_0$ is absolutely continuous, we obtain that $\PP(\xi=2k)=0$, and
 \begin{align*}
   \PP(\xi=2k+1)
    &= \PP\left( \{ h_0>0.1\}\cap \bigcap_{j=1}^k \Big\{ \frac{v_j^2}{2g} > 0.1 \Big\}  \cap \Big\{ \frac{ v_{k+1}^2}{2g} < 0.1 \Big\} \right)\\
    &= \PP\left( \{ h_0>0.1\} \cap \bigcap_{j=1}^k \Big\{ \frac{ (0.95^j\cdot v_0)^2}{2g} > 0.1 \Big\} \cap \Big\{ \frac{ (0.95^{k+1}\cdot v_0)^2}{2g} < 0.1 \Big\} \right) \\
    &= \PP\left( \{ h_0>0.1 \} \cap  \bigcap_{j=1}^k \Big\{ \frac{ 0.95^{2j}\cdot 2gh_0}{2g} > 0.1 \Big\} \cap \Big\{ \frac{ 0.95^{2(k+1)}\cdot 2g h_0}{2g} < 0.1 \Big\} \right)\\
    &= \PP\left( \frac{0.1}{0.95^{2k}} < h_0 < \frac{0.1}{0.95^{2(k+1)}}  \right)\\
    &=\begin{cases}
        \frac{0.1}{0.95^{2k}}\left( \frac{1}{0.95^2} - 1 \right)  & \text{if \ $\frac{0.1}{0.95^{2(k+1)}} < 1$,}\\
        1- \frac{0.1}{0.95^{2k}} & \text{if \ $\frac{0.1}{0.95^{2k}} < 1 < \frac{0.1}{0.95^{2(k+1)}}$,}\\
        0 & \text{if \ $  \frac{0.1}{0.95^{2k}} > 1$.}
      \end{cases}
 \end{align*}
Using that the inequality $\frac{0.1}{0.95^{2(k+1)}}< 1$ holds if and only if
 \begin{align*}
  \ln(0.1) < 2(k+1)\ln(0.95)
    \quad \Longleftrightarrow \quad
    \frac{\ln(0.1)}{2\ln(0.95)} - 1 > k
    \quad \Longleftrightarrow \quad  k\leq 21,
 \end{align*}
  since $k$ is an integer, and that the inequality $\frac{0.1}{0.95^{2k}} < 1$ holds if and only if
  \[
    \frac{\ln(0.1)}{2\ln(0.95)}  > k
    \quad \Longleftrightarrow \quad  k\leq 22,
  \]
 we get that for each $k\in\{1,2,\ldots\}$,
  \begin{align*}
  \PP(\xi=2k+1)
   = \begin{cases}
        0.1\left( \frac{1}{0.95^2} - 1 \right) 0.95^{-2k} & \text{if \ $k\in\{1,2,\ldots,21\}$,}\\
        1- \frac{0.1}{0.95^{2k}} & \text{if \ $k=22$,}\\
        0 & \text{if \ $k\geq 23$.}
      \end{cases}
  \end{align*}
All in all, the distribution of $\xi$ takes the form
 \begin{align*}
   &\PP(\xi=0) = 0.1,\\
   &\PP(\xi=2k+1) = 0.1\left( \frac{1}{0.95^2} - 1 \right) 0.95^{-2k} = 0.95^{-2k} \PP(\xi=1), \qquad k\in\{0,1,2,\ldots,21\},\\
   & \PP(\xi=45) = 1 - \frac{0.1}{0.95^{44}} \approx 0.04465.
 \end{align*}
One can check that the given values are indeed a probability distribution, since the given values are trivially nonnegative and
 \begin{align*}
  & 0.1 + \sum_{k=0}^{21} 0.1\left( \frac{1}{0.95^2} - 1 \right) 0.95^{-2k} + 1 - \frac{0.1}{0.95^{44}}\\
  &\qquad = 0.1 + 0.1\left( \frac{1}{0.95^2} - 1 \right) \cdot \frac{ (0.95^{-2})^{22} -1 }{ 0.95^{-2} -1 }
            +  1 - \frac{0.1}{0.95^{44}}
   =1.
 \end{align*}

This implies that the expectation of $\xi$ is
   \begin{align*}
     \EE(\xi)& = 0\cdot 0.1 + \sum_{k=0}^{21} (2k+1) \cdot 0.1\left( \frac{1}{0.95^2} - 1 \right) 0.95^{-2k}
               + 45\left(1 - \frac{0.1}{0.95^{44}}\right) \\
             & = 44.9 - \frac{4.4}{0.95^{44}} + 0.2\frac{1-0.95^2}{0.95^2} \sum_{k=1}^{21} k\cdot0.95^{-2k}
              \approx 27.35.
   \end{align*}

Note that the function $\{0,1,\ldots,21\}\ni k\mapsto \PP(\xi=2k+1)$ is strictly increasing,
 however, the function $\{0,1,\ldots,22\} \ni k \mapsto \PP(\xi=2k+1)$ is not strictly increasing, since
 $\PP(\xi=43)\approx 0.09315$, which is greater than $\PP(\xi=45)\approx 0.04465$.
Remark also that the (unique) maximum of the probabilities $\PP(\xi=0)$ and $\PP(\xi=2k+1)$, $k\in\{0,1,\ldots,22\}$ is equal to
 $\PP(\xi=0)=0.1$, yielding that the mode of the distribution of $\xi$ is $0$.
\proofend

The next exercise is a randomized version of Exercise 2, Round I, Category I
 at the S\'andor Mikola Physics Competition (2023) (see \cite{Mik}) in the sense that we
 consider the so-called collision number appearing in the exercise  as a random variable.
We also point out that the deterministic counterparts of the questions that we consider in our next exercise
 are not included in the exercise at the S\'andor Mikola Physics Competition.

\begin{Exe} \label{Ex_proj_motion_4}
According to the International Tennis Federation, a tennis ball is considered standard if,
 when dropped on concrete from a height of 254 cm, the rebound height falls between 135 cm and 147 cm.
By collision number, by definition, we mean the ratio of the rebound and landing speeds when a tennis ball
 is dropped on concrete from a height of 254 cm.
The companies A and B produce tennis balls.
In case of Company A, the collision number is a random variable with uniform distribution
 on the interval $[0.7, 0.8]$, and in case of Company B, it an absolutely continuous
 random variable with a density function
 \[
 f_B(x)= \begin{cases}
            200x-140 & \text{if $x\in[0.7, 0.8]$,}\\
            0 & \text{otherwise.}
         \end{cases}
 \]

\begin{enumerate}
 \item[(i)] We drop a tennis ball produced by Company A on concrete from a height of 254 cm.
           How high is the bounce back expected to be?
           Answer the same question for a tennis ball produced by Company B.
 \item[(ii)] What is the probability that a tennis ball produced by Company A is standard?
             Answer the same question for a tennis ball produced by Company B.
 \item[(iii)] Company A and B send us 40 and 60 tennis balls, respectively, as a promotional gift.
              We mix the 100 tennis balls, we randomly choose one of them and then, by dropping it
              on concrete from a height of 254 cm, we classify it standard.
              What is the probability that the tennis ball was produced by company A?
\end{enumerate}
\end{Exe}

\noindent{\bf Solution.}
If we drop a ball from a height of $h_l$, then it has velocity $v_l=\sqrt{2gh_l}$ when it hits the ground.
Furthermore, if its rebounce velocity is $v_f$, then, for the rebounce height denoted by $h_f$,
 it holds that $v_f=\sqrt{2gh_f}$.

With these notations, the collision number takes the form
  \[
    \frac{v_f}{v_l} = \frac{\sqrt{2gh_f}}{\sqrt{2g\cdot 2.54}} = \sqrt{\frac{h_f}{2.54}}.
  \]
In what follows, we denote the collision number by  $\xi$, i.e., $\xi:=\frac{v_f}{v_l}$.
Then the rebounce height is $h_f=2.54\cdot \xi^2$.

(i). In case of Company A, $\xi$ is uniformly distributed on the interval $[0.7, 0.8]$, and hence we have that
 \begin{align*}
   \EE(h_f) = 2.54 \EE(\xi^2) = 2.54 \int_{0.7}^{0.8} x^2 \frac{1}{0.1}\,\dd x
            = \frac{25.4}{3}(0.8^3-0.7^3)\approx 1.431,
 \end{align*}
 i.e., a tennis ball produced by Company A rebounds to the expected height of approximately 143.1 cm
 dropped from a height of 254 cm.

In case of Company B, $\xi$ is an absolutely continuous random variable with a density function $f_B$.
Then we get that
 \begin{align*}
   \EE(h_f)& = 2.54 \EE(\xi^2) = 2.54 \int_{0.7}^{0.8} x^2 (200x-140)\,\dd x
            = 2.54 \int_{0.7}^{0.8} (200x^3-140x^2)\,\dd x\\
           & = 2.54\Big( 50(0.8^4-0.7^4) - \frac{140}{3}(0.8^3-0.7^3)\Big) \approx 1.494,
 \end{align*}
 i.e., a tennis ball produced by Company B rebounds to the expected height of approximately 149.4 cm
 dropped from a height of 254 cm.

(ii).
Using that
 \[
    \Big\{  \text{a tennis ball is standard} \Big\} = \{ h_f\in(1.35, 1.47)\},
 \]
  we get that, in case of Company A, the probability that a tennis ball is standard is
 \begin{align*}
  \PP(h_f\in(1.35, 1.47))
     &= \PP( 1.35 < 2.54\cdot\xi^2 < 1.47)
     = \PP\Big( \sqrt{\frac{135}{254}} < \xi < \sqrt{\frac{147}{254}} \Big)\\
     &= \frac{\sqrt{\frac{147}{254}} - \sqrt{\frac{135}{254}}}{0.8 - 0.7}
     \approx 0.3171,
 \end{align*}
 and, in case of Company B, is
 \begin{align*}
  \PP(h_f\in(1.35, 1.47))
     &= \PP( 1.35 < 2.54\cdot\xi^2 < 1.47)
      = \PP\Big( \sqrt{\frac{135}{254}} < \xi < \sqrt{\frac{147}{254}} \Big)\\
     &=\int_{\sqrt{\frac{135}{254}}}^{\sqrt{\frac{147}{254}}} (200x-140)\,\dd x
      =\frac{1200}{254} - 140 \left( \sqrt{\frac{147}{254}} - \sqrt{\frac{135}{254}} \right)
     \approx 0.2847.
 \end{align*}

(iii).
Let us introduce the following events
 \begin{align*}
   &G_A:=\Big\{  \text{the chosen tennis ball was produced by Company A}  \Big\},\\
   &G_B:=\Big\{  \text{the chosen tennis ball was produced by Company B}  \Big\},\\
   &S:=\Big\{  \text{the chosen tennis ball is standard}  \Big\}.
 \end{align*}
We need to calculate the conditional probability $\PP(G_A\mid S)$.
Since $G_A$ and $G_B$ are disjoint events with positive probability and their union is the whole sample space,
 one can apply Bayes' theorem (see \eqref{Bayes}) and we get that
 \begin{align*}
  \PP(G_A\mid S) & = \frac{\PP(S\mid G_A)\PP(G_A)}{\PP(S\mid G_A)\PP(G_A) + \PP(S\mid G_B)\PP(G_B)} \\
                 & = \frac{\frac{\sqrt{\frac{147}{254}} - \sqrt{\frac{135}{254}}}{0.8 - 0.7} \cdot 0.4}
                        {\frac{\sqrt{\frac{147}{254}} - \sqrt{\frac{135}{254}}}{0.8 - 0.7} \cdot 0.4
                           + \left( \frac{1200}{254} - 140 \left( \sqrt{\frac{147}{254}} - \sqrt{\frac{135}{254}} \right) \right)\cdot 0.6} \\
                 & = \frac{4\left( \sqrt{\frac{147}{254}} - \sqrt{\frac{135}{254}} \right)}
                           {\frac{720}{254} - 80\left( \sqrt{\frac{147}{254}} - \sqrt{\frac{135}{254}} \right)}
                 \approx 0.4261.
 \end{align*}
\proofend

Part (i) of the next Exercise \ref{Ex_proj_motion_5} is a randomized version of Exercise 1, Round II, Category III
 at the S\'andor Mikola Physics Competition (2023) (see \cite{Mik}) in the sense that
 the precise theoretical value of the gravity of Earth is considered to be a random value.
Note that this kind of normalization is different from the ones in Exercises \ref{Ex_proj_motion_1}-\ref{Ex_proj_motion_3}
 in the sense that the uncertainty in a system's parameter is studied in Exercise \ref{Ex_proj_motion_5}
 instead of the uncertainty in some of the inputs of a dynamical system.

\begin{Exe}\label{Ex_proj_motion_5}
Let $g$ denote the (theoretical) value of the gravity of Earth and let $\xi$ be a uniformly distributed random variable on the interval $[-1,1]$.
What is the expectation of the relative error that we make if the gravity of Earth is taken to be $(g+\xi)$ instead of the theoretical value $g$ and
 \begin{itemize}
    \item[(i)] we calculate the distance traveled by a free-falling body over a given period of time?
    \item[(ii)] we calculate the time needed by a free-falling body to travel a given distance?
 \end{itemize}
\end{Exe}

\noindent{\bf Solution.}
It is known that a free-falling body travels a distance $\frac{g}{2}t^2$ over a given period of time $t$, and
 it needs a time $\sqrt{\frac{2h}{g}}$ to travel a given distance $h$.

(i). The relative error of the distances over a given time period $t$
  (with respect to the distance calculated by the theoretical value of $g$) is
 \[
     \frac{\frac{g+\xi}{2}t^2 - \frac{g}{2}t^2 }{ \frac{g}{2}t^2}
        = \frac{\xi}{g}.
 \]
Consequently, the expected value of $\frac{\xi}{g}$ is
 \[
  \EE\left(\frac{\xi}{g}\right)
      = \frac{1}{g}\EE(\xi)
      = \frac{1}{g}\cdot \frac{-1+1}{2}=0,
 \]
 that is, the expected value of the relative error in question is $0$.

(ii). The relative error of the times needed for travelling a given distance $h$
 (with respect to the time calculated by the theoretical value of $g$) is
 \begin{align*}
  \frac{ \sqrt{\frac{2h}{g+\xi}} - \sqrt{\frac{2h}{g}} }{ \sqrt{\frac{2h}{g}}}
        = \sqrt{\frac{g}{g+\xi}} - 1.
 \end{align*}
Consequently, we get that
 \begin{align*}
   &\EE\left( \sqrt{\frac{g}{g+\xi}} - 1\right)
     = \sqrt{g} \EE\left(\frac{1}{\sqrt{g+\xi}} \right) - 1
      = \sqrt{g} \int_{-1}^1 \frac{1}{\sqrt{g+x}}\cdot \frac{1}{2}\,\dd x - 1\\
    & = \sqrt{g}(\sqrt{g+1} - \sqrt{g-1})-1
     = \frac{2\sqrt{g}}{\sqrt{g+1} + \sqrt{g-1}} - 1
      =   \frac{2\sqrt{g} - \sqrt{g+1} - \sqrt{g-1} }{\sqrt{g+1} + \sqrt{g-1}}.
 \end{align*}
This expected value is positive, since the function
 $[0,\infty) \ni x  \mapsto \sqrt{x}$ is strictly concave, and hence
 \[
   \frac{\sqrt{g+1} + \sqrt{g-1}}{2} > \sqrt{\frac{1}{2}(g+1) + \frac{1}{2}(g-1)} = \sqrt{g}.
 \]
In our present paper, the theoretical value of $g$ is considered to be $9.81$,
 and hence the expected value of the relative error in question is approximately $0.001305$.

Finally, we call the attention to the fact that, in part (i), the expected value of the relative error
 does not depend on the theoretical value of $g$, while in part (ii), it does.
\proofend

Regarding didactic conclusions, we can say that our exercises are suitable 
 for introducing some basic concepts of probability theory, such as expected value, 
 variance and density function of a random variable, in a physical context. 
They enable comparison between stochastic and deterministic solutions 
 (for example, in case of part (i) of Exercise \ref{Ex_proj_motion_1}, see part (i) of Remark \ref{Rem_Exercise21}).
Exercise \ref{Ex_proj_motion_4} encourages the exploration of different probability distributions 
 to describe the behavior of a variable system.
Moreover, students can extend the method to similar systems to develop their problem-solving skills 
(for example, for some natural counterparts of Exercise \ref{Ex_proj_motion_1}, see part (ii) of Remark \ref{Rem_Exercise21}).

\section{A stochastic approach in statics}\label{section_statics}

In this section, we demonstrate our method for incorporating a stochastic point of view
 into physics exercises of mathematics education by presenting exercises for statics.
In them, we randomize the geometry of the static problem, for example,
 the shape of a plate or the position of some of the pinning points of a plate.
To avoid technicalities, we choose uniform distributions as the distributions of the random inputs.
Altogether, we present three exercises in this section.
These exercises highlight the importance of probabilistic tools 
 in static problems, they can make abstract concepts of probability theory more tangible.

\begin{figure}[ht]
 \centering
 \includegraphics[height=7cm]{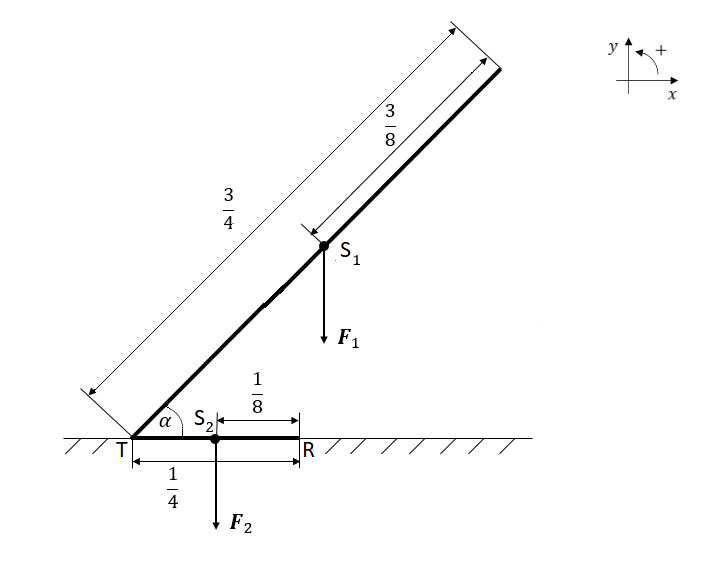}
  \caption{a planar section of the bent plate for Exercise \ref{Ex_statics_1} and its solution.}
 \label{Fig3_static}
\end{figure}

\begin{Exe}\label{Ex_statics_1}
A square piece of homogeneous sheet metal of side length 1 m and mass $m$ kg is bent in the following way:
 on one side, at a point chosen at a distance of $\frac{1}{4}$ from one of the vertices, a perpendicular is set on that side,
 and we bend the metal sheet along this perpendicular in a way that the angle between the two pieces is uniformly distributed
 on the interval $(0,\frac{\pi}{2})$.
We place the bent plate with its smaller part on a horizontal surface (see Figure \ref{Fig3_static}), and then release it.
\begin{itemize}
  \item[(i)] What is the probability that it does not fall over?
   \item[(ii)] What is the probability that the perpendicular projection of the
               center of gravity of the larger part of the bent plate
               falls on the smaller part of the bent plate given that the bent plate does not fall over?
\end{itemize}
\end{Exe}

\noindent{\bf Solution.}
 Figure \ref{Fig3_static} shows a planar section of the bent plate corresponding to a plane,
 which is perpendicular to the bending line at a point (denoted by $T$).
Let  $\alpha$ denote the angle between the two pieces.
Then $\alpha$ is a random variable with uniform distribution on the interval $(0,\frac{\pi}{2})$.
Let $S_1$ and $S_2$ be the centers of gravity of the larger and smaller parts of the plate, respectively,
 and let $R$ be the endpoint different from $T$ of the planar section of the smaller part.
Let
 \[
    \bF_1:= \begin{bmatrix}
              0 \\
              -mg\frac{3}{4} \\
            \end{bmatrix}
        \qquad \text{and} \qquad
    \bF_2:= \begin{bmatrix}
              0 \\
              -mg\frac{1}{4} \\
            \end{bmatrix}.
\]

(i).
Then $\{\text{the bent plate does not fall over}\} = A\cup B$, where the events $A$ and $B$ are defined by
 \[
  A:=\Big\{ \text{the perpendicular projection of $S_1$ falls on the section $TR$} \Big\},
 \]
 and
 \begin{align*}
   B:= &\Big\{ \text{the perpendicular projection of $S_1$ does not fall on the section $TR$} \Big\} \\
       &\cap \Big\{ \text{the magnitude of the moment of force of $\bF_2$ with respect to $R$ is greater than }\\
       &\phantom{\cap \Big\{ \Big\{ } \text{or equal to the magnitude of the moment of force of $\bF_1$ with respect to $R$} \Big\}.
 \end{align*}
Then the events $A$ and $B$ are disjoint, and using that the distance between the perpendicular projection of $S_1$ and the point $T$ is
 $\frac{3}{8}\cos(\alpha)$ and the length of the section $TR$ is $\frac{1}{4}$, we get that
 \begin{align*}
   &A=\left\{ \frac{3}{8}\cos(\alpha)\leq \frac{1}{4}\right\},\\
   &B=\left\{ \frac{3}{8}\cos(\alpha)> \frac{1}{4} \right\}
       \cap \left\{  mg\frac{1}{4}\cdot \frac{1}{8} \geq mg\frac{3}{4}\cdot\left(\frac{3}{8}\cos(\alpha) - \frac{1}{4}\right) \right\}.
 \end{align*}
Consequently, we have that $A=\Big\{\cos(\alpha)\leq \frac{2}{3}\Big\}$ and
 \begin{align*}
  B & = \left\{ \cos(\alpha)> \frac{2}{3} \right\}
       \cap \left\{ \frac{1}{24} \geq  \frac{3}{8}\cos(\alpha) - \frac{1}{4} \right\}
      = \left\{ \cos(\alpha)  > \frac{2}{3} \right\}
        \cap \left\{ \cos(\alpha) \leq \frac{7}{9} \right\} \\
    & = \left\{ \frac{2}{3} < \cos(\alpha) \leq \frac{7}{9} \right\} .
 \end{align*}
Hence $\{\text{the bent plate does not fall over}\} = \{   \cos(\alpha) \leq \frac{7}{9}\}$.
Here the inequality $\cos(\alpha) \leq \frac{7}{9}$, $\alpha\in(0,\frac{\pi}{2})$ holds if and only if
 $\alpha\in[\alpha_0,\frac{\pi}{2})$, where $\alpha_0\in (0,\frac{\pi}{2})$ is the unique value for which
 $\cos(\alpha_0)=\frac{7}{9}$, i.e., $\alpha_0=\arccos\left(\frac{7}{9}\right)\approx 0.6797$.
Consequently, we have
 \[
   \{\text{the bent plate does not fall over}\} = A\cup B = \left\{ \alpha_0 \leq \alpha < \frac{\pi}{2} \right\},
 \]
 and note that the angle $\alpha_0$ is the minimum of those angles in $(0,\frac{\pi}{2})$ for which
 the bent plate is in equilibrium, and it holds that
 \[
    mg\frac{1}{4}\cdot \frac{1}{8} = mg\frac{3}{4}\cdot\left(\frac{3}{8}\cos(\alpha_0) - \frac{1}{4}\right).
 \]
Therefore, we get that
 \begin{align*}
  &\PP(\{\text{the bent plate does not fall over}\})
      = \PP\left(\alpha_0 \leq \alpha < \frac{\pi}{2}\right) \\
 &\qquad \qquad = \frac{\frac{\pi}{2} - \alpha_0}{\frac{\pi}{2}}
       = \frac{\frac{\pi}{2} - \arccos\left(\frac{7}{9}\right)}{\frac{\pi}{2}}
      = 1 - \frac{2\arccos\left(\frac{7}{9}\right)}{\pi}
      \approx 0.5673.
 \end{align*}

(ii).
Using the notations introduced in the solution of part (i), we need to calculate the conditional probability
 $\PP(A\mid A\cup B)$.
Since the events $A$ and $B$ are disjoint, we get that
 \[
   \PP(A\mid A\cup B) = \frac{\PP(A\cap(A\cup B))}{\PP(A\cup B)}
                      = \frac{\PP(A)}{\PP(A\cup B)},
 \]
 where, by the solution of part (i),
 \[
    \PP(A\cup B) = 1 - \frac{2\arccos\left(\frac{7}{9}\right)}{\pi}.
 \]
Furthermore, we have that
 \[
   \PP(A) = \PP\left(\frac{3}{8}\cos(\alpha)\leq \frac{1}{4}\right)
          =  \PP\left(\cos(\alpha)\leq \frac{2}{3}\right).
 \]
Here the inequality $\cos(\alpha) \leq \frac{2}{3}$, $\alpha\in(0,\frac{\pi}{2})$ holds if and only if
 $\alpha\in[\alpha_1,\frac{\pi}{2})$, where $\alpha_1\in (0,\frac{\pi}{2})$ is the unique value for which
 $\cos(\alpha_1)=\frac{2}{3}$, i.e., $\alpha_1=\arccos\left(\frac{2}{3}\right)\approx 0.8411$.
Consequently, we obtain that
 \[
  \PP(A) = \PP\left(\alpha_1 \leq \alpha < \frac{\pi}{2}\right)
         = \frac{\frac{\pi}{2} - \alpha_1}{\frac{\pi}{2}}
       = \frac{\frac{\pi}{2} - \arccos\left(\frac{2}{3}\right)}{\frac{\pi}{2}}
      = 1 - \frac{2\arccos\left(\frac{2}{3}\right)}{\pi}
      \approx 0.4645.
 \]
Then we have
 \begin{align*}
   \PP(A\mid A\cup B) = \frac{\frac{\pi}{2} - \arccos\left(\frac{2}{3}\right) }{\frac{\pi}{2} - \arccos\left(\frac{7}{9}\right)}
                      \approx 0.8189.
 \end{align*}
\proofend

The following exercise is a randomized version of Problem 2 in Section 7.5 in Sz\'\i ki \cite{Szi}.

\begin{figure}[ht]
 \centering
 \includegraphics[width=8cm]{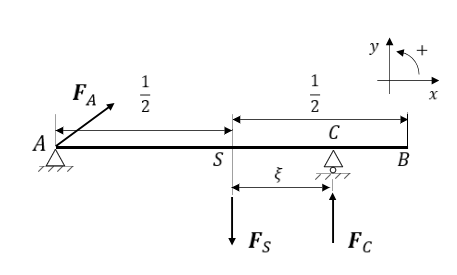}
  \caption{a prismatic beam fixed by means of a pin support at point $A$
            and a roller support at point $C$ for Exercise \ref{Ex_statics_2} and its solution.}
 \label{Fig4_static}
\end{figure}

\begin{Exe}\label{Ex_statics_2}
Let us consider a homogeneous prismatic beam of length 1 m and mass $m$ kg, and its two endpoints and the centre point are denoted by
 $A$, $B$ and $S$, respectively.
The beam is fixed by means of a pin support at its point $A$, and of a roller support at a point chosen uniformly
 on the section $SB$ (see Figure \ref{Fig4_static}).
 \begin{itemize}
  \item[(i)] Let us determine the expected values of the reaction forces at the two supporting points!
   \item[(ii)] Are the expected values of the reaction forces calculated in part (i) equal to the
              reaction forces which correspond to the case when the beam is fixed by means of a pin support at its point $A$,
              and of a roller support at the bisection point of the section $SB$?
              Why is there no contradiction compared to the result obtained in part (i)?
\end{itemize}
\end{Exe}

\noindent{\bf Solution.}
(i). Let $C$ and $\xi$ denote the point corresponding to the roller support and the length of the section $SC$, respectively.
Due to the assumptions, $\xi$ is a random variable having uniform distribution on the interval $[0,\frac{1}{2}]$.
The reaction forces at the points $A$ and $C$ are denoted by $\bF_A=[F_{A,x},F_{A,y}]^\top$ and $\bF_C=[F_{C,x},F_{C,y}]^\top$,
 respectively.
Furthermore, let $\bF_S:=[0,-mg]^\top$.
Note also that the direction of rotation is positive (i.e., counterclockwise) on Figure \ref{Fig4_static}.

The beam is in equilibrium if and only if the resultant force is zero and the resultant moment of the force system
 that acts on the beam is equal to zero for a given point of the plane.
Consequently, the beam is in equilibrium if and only if the resultant force is
 \begin{align}\label{help2}
    \bF_A + \bF_C + \bF_S = \bzero,
 \end{align}
 and the resultant moment of forces with respect to the point $A$ is
 \begin{align}\label{help3}
    -mg\cdot\frac{1}{2} + F_{C,y}\left(\frac{1}{2}+\xi\right) = 0.
 \end{align}
Using that the point $C$ corresponds to a roller support, we have that $F_{C,x}=0$, and hence, writing the vectors in the equation
  \eqref{help2} in coordinate form, we get that
 \begin{align}\label{help4}
   \begin{bmatrix}
     F_{A,x} \\
     F_{A,y} \\
   \end{bmatrix}
   + \begin{bmatrix}
       0 \\
     F_{C,y} \\
   \end{bmatrix}
   + \begin{bmatrix}
     0 \\
     -mg \\
   \end{bmatrix}
  =\begin{bmatrix}
     0 \\
     0 \\
   \end{bmatrix}.
 \end{align}
This shows that $F_{A,x}=0$ and $F_{A,y} + F_{C,y} = mg$.
Further, by \eqref{help3}, we obtain that
 \[
  F_{C,y} = \frac{mg}{2\left(\frac{1}{2}+\xi\right)}
          = mg\cdot \frac{1}{2\xi+1},
 \]
 and hence
 \[
   F_{A,y} = mg - F_{C,y} = mg - \frac{mg}{2\xi+1}
           = mg \cdot\frac{2\xi}{2\xi+1}.
 \]
Consequently, we have that
 \begin{align}\label{help_stat1}
   \bF_A = \begin{bmatrix}
             0 \\
             mg\cdot\frac{2\xi}{2\xi+1} \\
           \end{bmatrix}
           \qquad \text{and} \qquad
   \bF_C = \begin{bmatrix}
             0 \\
             mg\cdot\frac{1}{2\xi+1} \\
           \end{bmatrix}.
 \end{align}
Therefore, the expected values of the reaction forces at the points $A$ and $C$ take the following forms:
 \[
  \EE(\bF_A) = \begin{bmatrix}
                  0 \\
                 mg\EE\left(\frac{2\xi}{2\xi+1}\right) \\
               \end{bmatrix}
             = \begin{bmatrix}
                  0 \\
                 mg(1-\ln(2)) \\
               \end{bmatrix},
 \]
 and
  \[
  \EE(\bF_C) = \begin{bmatrix}
                  0 \\
                 mg\EE\left(\frac{1}{2\xi+1}\right) \\
               \end{bmatrix}
             = \begin{bmatrix}
                  0 \\
                 mg\ln(2) \\
               \end{bmatrix},
 \]
 since
 \begin{align*}
 &\EE\left(\frac{1}{2\xi+1}\right)
   = \int_0^{1/2} \frac{1}{2x+1}\cdot 2 \,\dd x
   = \ln(2) - \ln(1) = \ln(2),\\
 &\EE\left(\frac{2\xi}{2\xi+1}\right)
   = 1 - \EE\left(\frac{1}{2\xi+1}\right)
   = 1 - \ln(2),
 \end{align*}
 where we used Newton-Leibniz formula and that a primitive function of the function $[0,\frac{1}{2}]\ni x\mapsto 2/(2x+1)$ is
 $[0,\frac{1}{2}]\ni x\mapsto\ln(2x+1)$.

(ii).
Similarly to the solution of part (i) (formally replacing $\xi$ by $\frac{1}{4}$ in the formula \eqref{help_stat1}),
 we get that
 \[
   \bF_A = \begin{bmatrix}
             0 \\
             \frac{mg}{3} \\
           \end{bmatrix}
           \qquad \text{and} \qquad
   \bF_C = \begin{bmatrix}
             0 \\
             \frac{2mg}{3}\\
           \end{bmatrix}.
 \]
Since $\frac{1}{3}\ne 1-\ln(2)$ and $\frac{2}{3}\ne \ln(2)$, the reaction forces above are not equal to
 the expected reaction forces calculated in the solution of part (ii).
This is not a contradiction, since $\EE(\frac{1}{2\xi+1}) \ne \frac{1}{\EE(2\xi+1)} = \frac{1}{2\cdot\frac{1}{4}+1} = \frac{2}{3}$.
\proofend

\begin{Exe}\label{Ex_statics_3}
The plate with the dimensions given in Figure \ref{Fig1_Static3} is fixed by means of a pin support at point $A$ and a rope at point $B$,
 and is subjected to concentrated forces $\bF_1$ and $\bF_2$, respectively.
Assume that the mass of the plate can be neglected,
 and the magnitude of the forces $\bF_1$ and $\bF_2$ is $300\,$N and $100\,$N, respectively.
Further, we suppose that the distribution of $b$ is uniform distribution on the interval $(0 \,\mathrm{m},\sqrt{3} \,\mathrm{m})$.
 \begin{figure}[ht]
 \centering
 \includegraphics[height=5cm]{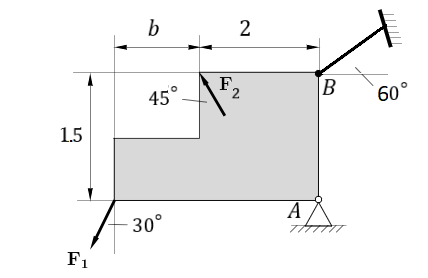}
  \caption{a plate fixed by means of a pin support at point $A$ and a rope at point $B$,
            and is subjected to concentrated forces $\bF_1$ and $\bF_2$, respectively for Exercise \ref{Ex_statics_3} and its solution.}
 \label{Fig1_Static3}
 \end{figure}
 \begin{itemize}
   \item[(i)] What is the distribution of the rope force at point $B$?
   \item[(ii)] What is the probability that the magnitude of the rope force at point $B$ is in the interval $\left[400\sqrt{3}\,\mathrm{N}, 400\sqrt{3}+100 \,\mathrm{N}\right]$?
   \item[(iii)] What is the expected value of the magnitude of the rope force at point $B$?
   \item[(iv)] What is the expected value of the force at point $A$?
 \end{itemize}
\end{Exe}

\noindent{\bf Solution.}
Let us denote by $\bF_A$ and $\bF_B$ the forces at the points $A$ and $B$, respectively.
In part (i) we need to determine the distribution of $\Vert \bF_B\Vert$;
 in part (ii) the probability $\PP(\Vert \bF_B\Vert\in(400\sqrt{3},400\sqrt{3}+100))$;
 in part (iii) the expected value of $\Vert \bF_B\Vert$; and in part (iv) the expected value of $\bF_A$.

Recall that the coordinates of a vector $\bv$ in the $(x,y)$-plane are denoted by $v_x$ and $v_y$, respectively, i.e., $\bv=(v_x,v_y)$.
 Accordingly, we have $\bF_1=[F_{1,x},F_{1,y}]^\top$, $\bF_2=[F_{2,x},F_{2,y}]^\top$, $\bF_A=[F_{A,x},F_{A,y}]^\top$ and $\bF_B=[F_{B,x},F_{B,y}]^\top$.
We have $\Vert \bF_1\Vert=300$, $\Vert \bF_2\Vert=100$, and
 \begin{align*}
    &F_{1,x} = - \Vert \bF_1\Vert \cdot \sin\left(\frac{30}{360}\cdot2\pi\right) = -300\cdot \frac{1}{2} = -150,\\
    &F_{1,y} = - \Vert \bF_1\Vert \cdot  \cos\left(\frac{30}{360}\cdot2\pi\right) = -300\cdot \frac{\sqrt{3}}{2} = -150\sqrt{3},
\end{align*}
\begin{align*}    
    &F_{2,x} = - \Vert \bF_2\Vert \cdot  \cos\left(\frac{45}{360}\cdot2\pi\right) = -100\cdot \frac{\sqrt{2}}{2} = -50\sqrt{2},\\
    &F_{2,y} =  \Vert \bF_2\Vert \cdot  \sin\left(\frac{45}{360}\cdot2\pi\right) = 100\cdot \frac{\sqrt{2}}{2} = 50\sqrt{2},\\
    &F_{B,x} =  \Vert \bF_B\Vert \cdot  \cos\left(\frac{60}{360}\cdot2\pi\right) = \Vert \bF_B\Vert \cdot\frac{1}{2},\\
    &F_{B,y} =  \Vert \bF_B\Vert \cdot  \sin\left(\frac{30}{360}\cdot2\pi\right) = \Vert \bF_B\Vert \cdot\frac{\sqrt{3}}{2}.
 \end{align*}
We take the direction of rotation as positive (i.e., counterclockwise) on Figure \ref{Fig1_Static3}.
Hence the moments $\underset{A}{M_1}$, $\underset{A}{M_2}$ and $\underset{A}{M_B}$ of the forces $\bF_1$, $\bF_2$ and $\bF_B$
 with respect to the point $A$ are
 \begin{align*}
   &\underset{A}{M_1} := \Vert \bF_1\Vert \cdot  \cos\left(\frac{30}{360}\cdot2\pi\right)\cdot (2+b)
                      = 300\frac{\sqrt{3}}{2}(2+b)
                      = 150\sqrt{3}(2+b),\\
   &\underset{A}{M_2} := \Vert \bF_2\Vert \cdot  \cos\left(\frac{45}{360}\right)\cdot 1.5
                        -\Vert \bF_2\Vert \cdot  \sin\left(\frac{45}{360}\right)\cdot 2
                      = 150\frac{\sqrt{2}}{2} - 100\sqrt{2}
                      = -25\sqrt{2},
\end{align*}                      
 and                     
\begin{align*}                      
  &\underset{A}{M_B} := - \Vert \bF_B\Vert \cdot \cos\left(\frac{60}{360}\cdot2\pi\right)\cdot 1.5
                     = - \Vert \bF_B\Vert \cdot  \frac{1}{2}\cdot 1.5 = -\frac{3}{4} \Vert \bF_B\Vert.
 \end{align*}
Since the plate is balanced, the equilibrium of forces in the directions $x$ and $y$ and
 that of the moment of forces with respect to the point $A$ are as follows:
 \begin{align*}
  &F_{A,x} + F_{B,x} + F_{1,x} + F_{2,x} =0,\\
  &F_{A,y} + F_{B,y} + F_{1,y} + F_{2,y} =0,\\
  &\underset{A}{M_B}+\underset{A}{M_1}+\underset{A}{M_2} = 0.
 \end{align*}
This system of equations takes the form
 \begin{align*}
 	F_{A,x} + \frac12 \Vert \bF_B\Vert &= 150 + 50\sqrt{2},\\
 	F_{A,y} + \frac{\sqrt{3}}{2}\Vert \bF_B\Vert &= 150\sqrt{3} - 50\sqrt{2},\\
 	 - \frac{3}{4}\Vert \bF_B\Vert &= -150\sqrt{3}(2+b)+25\sqrt{2}.
 \end{align*}
This is a system of linear equations in the variables $F_{A,x}$, $F_{A,y}$ and $\Vert \bF_B\Vert$.
It follows from the last equation that
\[
\Vert \bF_B\Vert=400\sqrt{3}-\frac{100}{3}\sqrt{2}+200\sqrt{3}b.
\]
Substituting the value of $\Vert \bF_B\Vert$ to the second equation, we get that%
 \[
  F_{A,y}=150\sqrt{3}-50\sqrt{2}-600+50\frac{\sqrt{6}}{3}-300b.
 \]
By substituting the value of $\Vert \bF_B\Vert$ to the first equation, we have
 \[
 F_{A,x}=150+\frac{200}{3}\sqrt{2}-200\sqrt{3}-100\sqrt{3}b.
 \]

(i).
Since $b$ has a uniform distribution on the interval $(0,\sqrt{3})$, its  cumulative distribution function takes the form
 \[
   \PP( b <x)
      = \begin{cases}
           0 & \text{if $x\leq 0$},\\
           \frac{x}{\sqrt{3}} & \text{if $x\in(0,\sqrt{3})$},\\
           1 & \text{if $x\geq \sqrt{3}$}.
        \end{cases}
 \]
Hence the cumulative distribution function of $\Vert \bF_B\Vert$ can be written as
 \begin{align*}
 	\PP(\Vert \bF_B\Vert<x)
    & = \PP\left(400\sqrt{3}-\frac{100}{3}\sqrt{2}+200\sqrt{3}b<x\right)\\
 	& =\PP\left(b<\frac{x-400\sqrt{3}+\frac{100}{3}\sqrt{2}}{200\sqrt{3}}\right)\\ 
 	&= \begin{cases}
         0 & \text{if $x\leq 400\sqrt{3}+\frac{100}{3}\sqrt{2}$,}\\
         \frac{x - 400\sqrt{3}+\frac{100}{3}\sqrt{2}}{200\sqrt{3}} & \text{if $x\in \left(400\sqrt{3}-\frac{100}{3}\sqrt{2}, 400\sqrt{3}-\frac{100}{3}\sqrt{2}+600\right)$,}\\
         1 & \text{if $x\geq 400\sqrt{3}-\frac{100}{3}\sqrt{2}+600$.}
      \end{cases}
 \end{align*}
Consequently, $\Vert \bF_B\Vert$ has a uniform distribution on the interval
 \[
  \left(400\sqrt{3}-\frac{100}{3}\sqrt{2}, 400\sqrt{3}-\frac{100}{3}\sqrt{2}+600\right).
 \]

(ii). Using the results in the solution of part (i), we get that
 \begin{align*}
 &\PP\left(\Vert \bF_B\Vert \in \left[400\sqrt{3},400\sqrt{3}+100\right]\right)\\
 &=\frac{400\sqrt{3}+100-400\sqrt{3}+\frac{100}{3}\sqrt{2}}{600}
    - \frac{400\sqrt{3}-400\sqrt{3}+\frac{100}{3}\sqrt{2}}{600}
  = \frac16.
 \end{align*}
We note that this probability can be calculated using geometric probabilities as well.
Namely, the length of the interval $\left[400\sqrt{3},400\sqrt{3}+100\right]$ and
 $\left(400\sqrt{3}-\frac{100}{3}\sqrt{2}, 400\sqrt{3}-\frac{100}{3}\sqrt{2}+600\right)$ is $100$ and $600$, respectively,
 and hence the probability in question is $\frac{100}{600}=\frac{1}{6}$.

(iii). The expected value of $\Vert \bF_B\Vert$ is
 \[
 \dfrac{400\sqrt{3}-\frac{100}{3}\sqrt{2}+400\sqrt{3} - \frac{100}{3}\sqrt{2}+600}{2}
        =400\sqrt{3}-\frac{100}{3}\sqrt{2}+300
        \approx 945.7.
 \]

(iv).
The expected value of $\bF_A$ is
 \begin{align*}
  \EE(\bF_A)
    & = \begin{bmatrix}
         \EE(F_{A,x}) \\
         \EE(F_{A,y}) \\
       \end{bmatrix}
     = \begin{bmatrix}
         \EE\big(150+\frac{200}{3}\sqrt{2}-200\sqrt{3}-100\sqrt{3}b \big)\\
         \EE\big(150\sqrt{3}-50\sqrt{2}-600+50\frac{\sqrt{6}}{3}-300b\big) \\
       \end{bmatrix} \\
    & = \begin{bmatrix}
         150+\frac{200}{3}\sqrt{2}-200\sqrt{3}-100\sqrt{3}\EE(b)\\
         150\sqrt{3}-50\sqrt{2}-600+50\frac{\sqrt{6}}{3}-300\EE(b) \\
        \end{bmatrix}.
 \end{align*}
Using that $\EE(b)=\frac{\sqrt{3}}{2}$, we have that
 \begin{align*}
  \EE(\bF_A)
       = \begin{bmatrix}
           \frac{200}{3}\sqrt{2}-200\sqrt{3}\\
          -600-50\sqrt{2}+50\frac{\sqrt{6}}{3}\\
        \end{bmatrix}
      \approx
        \begin{bmatrix}
          -252.1\\
          -629.9\\
        \end{bmatrix}.
  \end{align*}
\proofend

Regarding didactic conclusions, 
  we can say that our exercises can draw the learners' attention 
  to the importance of the random positioning of the structural elements and its consequences 
  in static problems.

\section{Concluding remarks}

Our method for incorporating a stochastic point of view into physics exercises of mathematics education
 may be used by students and teachers familiar with elementary probability theory 
 and mechanics for understanding some basic concepts of stochastic mechanics.
The core of the method is the randomization of some inputs, the system model used does not differ
 from what we would use in the deterministic approach.
We demonstrate our method of randomization of some inputs by considering exercises from
the theory of projectile motion and statics.
 STEM students may be asked to create similar exercises, for other fields of physics as well.
Our stochastic perspective can help the students recognize the fact that many
 phenomena in our daily lives have a random nature, and therefore it can be useful to get in
 touch with stochastic thinking and tools in order to be able to understand these phenomena
 much better.
Computer simulations using Monte Carlo methods might also be assigned
 to our elaborate examples in order to confirm the solutions empirically.

\vspace*{5mm}

\appendix

\vspace*{5mm}

\noindent{\bf\Large Appendices}

\section{Basic knowledge of probability}\label{appProbab}

In this appendix, we recall some of the notations, conventions, concepts and results of probability theory that are used in solving the exercises.

In all of the paper, $(\Omega,\cA,\PP)$ denotes a probability space, where $\Omega$ is called sample space.
A mapping $\xi:\Omega\to\RR$ is called a random variable
 if $\{ \omega\in\Omega : \xi(\omega)<x\}\in\cA$ for all $x\in\RR$.
By a $2$-dimensional random vector, we mean a mapping $(\xi,\eta):\Omega\to\RR^2$, where $\xi$ and $\eta$
 are (1-dimensional) random variables.
Given a random variable $\xi:\Omega\to\RR$ and a real number $x\in\RR$,
 we use the notational conventions $\{\xi< x\} := \{ \omega\in\Omega : \xi(\omega)<x\}$,
 and accordingly $\PP(\xi<x):=\PP(\{\xi<x\})$.

Given events $A,B\in\cA$ such that $\PP(B)>0$, the conditional probability of $A$ given $B$ is defined by
 $\frac{\PP(A\cap B)}{\PP(B)}$, and is denoted by $\PP(A\mid B)$.
Given events $A,B\in\cA$ such that $\PP(A)\in(0,1)$ and $\PP(B)>0$, Bayes's theorem states that
 \begin{align}\label{Bayes}
   \PP(A\mid B) = \frac{\PP(B\mid A) \PP(A)}{\PP(B\mid A) \PP(A) + \PP(B\mid \Omega\setminus A) \PP( \Omega\setminus A)}.
 \end{align}

In some cases, we calculate the probability of an event using the method of geometric probabilities.
In this case, the underlying probability space is chosen as a (Borel) subset $S$ of
 $\RR^2$ furnished with the sigma-algebra $\cB(S)$ generated by all of the open subsets of $S$
 (i.e., with the Borel sigma-algebra of $S$), and, for all $A\in\cB(S)$,
 the probability $\PP(A)$ of $A$ is defined as the fraction of the area ($2$-dimensional Lebesgue measure)
 of $A$ and that of $S$, where we suppose that the area of $S$ is positive.

The cumulative distribution function of a random variable $\xi:\Omega\to\RR$ is defined by $F_\xi:\RR\to\RR$,
 $F_\xi(x):=\PP(\xi<x)$, $x\in\RR$.
A random variable is called discrete if its range $\xi(\Omega)$ is countable.
By the (probability) distribution of a discrete random variable $\xi$ having range $\xi(\Omega)=:\{x_i: i\in\NN\}$,
 we mean the sequence $(\PP(\xi=x_i))_{i\in\NN}$.
A random variable $\xi:\Omega\to\RR$ is called absolutely continuous if there exists a (Borel measurable) non-negative function $f_\xi:\RR\to\RR$
 such that $F_\xi(x)=\int_{-\infty}^x f_\xi(t)\,\dd t$, $x\in\RR$.
The expected value (expectation) of a discrete random variable $\xi$ having range $\xi(\Omega)=:\{x_i: i\in\NN\}$ is defined by
 $\EE(\xi):=\sum_{i=1}^\infty x_i \PP(\xi=x_i)$, provided that $\sum_{i=1}^\infty \vert x_i\vert \PP(\xi=x_i)<\infty$.
The expected value of a discrete random variable having finite range is nothing else but 
 the weighted average of the possible outcomes of the random variable, where the weights are the probabilities that the outcomes will occur.
The expected value (expectation) of an absolutely continuous random variable $\xi:\Omega\to\RR$ with a density function $f_\xi$
 is defined by $\EE(\xi):=\int_{-\infty}^\infty x f_\xi(x)\,\dd x$, provided that $\int_{-\infty}^\infty \vert x\vert  f_\xi(x)\,\dd x<\infty$.
The variance and standard deviation of a random variable $\xi$ is defined by $\DD^2(\xi):=\EE((\xi-\EE(\xi))^2)$ and $\DD(\xi):=\sqrt{\EE((\xi-\EE(\xi))^2)}$,
 respectively, provided that $\EE(\xi^2)<\infty$.
It is know that $\DD^2(\xi)=\EE(\xi^2) - (\EE(\xi))^2$.
The standard deviation of a random variable can be interpreted as a measure of the amount of 
 spread of the values of the random variable around its expected value.
A higher standard deviation indicates a greater spread, meaning the random variable is more likely to take on values far from its expected value.
The relative standard deviation (also known as coefficient of variation)
 of a positive random variable $\xi$ having finite second moment (i.e., $\EE(\xi^2)<\infty$) is given by $\rsd(\xi):=\DD(\xi)/\EE(\xi)$.
Roughly speaking, the higher relative standard deviation of a nonnegative random variable $\xi$,
 the greater the level of dispersion of the values of $\xi$ around its mean $\EE(\xi)$.
If $(\xi,\eta)$ is a $2$-dimensional random vector such that $\EE(\xi)$ and $\EE(\eta)$ are defined, then the expectation vector of $(\xi,\eta)$
 is given by $\EE(\xi,\eta):= (\EE(\xi), \EE(\eta))$.
If $\xi:\Omega\to\RR$ is a random variable such that $\EE(\xi)$ is defined and $a,b\in\RR$,
 then $\EE(a\xi + b)$ is defined as well and
 \begin{align}\label{exp_linearity}
   \EE(a\xi + b) = a\EE(\xi) + b.
 \end{align}

If $\xi, \eta:\Omega\to\RR$ are random variables, then the function
 $F_{\xi,\eta}:\RR^2\to\RR$, $F_{\xi,\eta}(x,y):=\PP(\xi<x,\eta<y)$, $x,y\in\RR$, is called the joint distribution function
 of $\xi$ and $\eta$.
The random variables $\xi,\eta:\Omega\to\RR$ are called independent if $F_{\xi,\eta}(x,y) = F_\xi(x)F_\eta(y)$, $x,y\in\RR$,
 i.e., the joint distribution function of $\xi$ and $\eta$ is the product of the distribution functions of $\xi$ and $\eta$.
If $\xi$ and $\eta$ are independent random variables and $g,h:\RR\to\RR$ are (Borel measurable) functions, then the random variables
 $g(\xi)$ and $h(\eta)$ are also independent.
If $\xi$ and $\eta$ are independent random variables such that $\EE(\xi)$ and $\EE(\eta)$ are defined, then
 $\EE(\xi\eta)$ is defined as well and
 \begin{align}\label{exp_fuggetlenseg}
  \EE(\xi\eta)=\EE(\xi)\EE(\eta).
 \end{align}

Finally, we formulate the law of total expectation that is used in the solution of Exercise \ref{Ex_proj_motion_1}.
Let $(\Omega,\mathcal{A},\PP)$ be a probability space, $\xi:\Omega\to\RR$ and $\eta:\Omega\to\RR$
 are independent random variables such that $\eta$ is absolutely continuous with a density function $f_\eta$.
Then for any (Borel measurable) function $G:\RR^2\to \RR$, we have that
 \begin{align}\label{telj_varh_ertek_tetele}
    \PP( G(\xi,\eta)<z) = \int_{-\infty}^\infty \PP( G(\xi,r)<z) f_\eta(r)\,\dd r, \qquad z\in\RR.
 \end{align}

\section{Basic knowledge of mechanics}\label{appPhys}

In this appendix, we provide some concepts of mechanics used in solving the exercises.

Motion in a homogeneous gravitational field (projectile motion) is discussed in the simplest planar model,
  assuming that the drag force (air resistance) is negligible.
In this case, the only force acting on a point-like object is the constant gravitational force
 ($m \cdot g$), and the equation of the motion is a parabola.
This type of motion is the superposition of a uniform motion in the horizontal direction
 and a uniformly accelerated motion (''free fall'') in the vertical direction.

In general, the initial velocity is arbitrary; its two components are the initial velocities in the horizontal and vertical directions.
In our examples, only one of the initial velocity components is different from zero.
Since the relevant formulas are well known and can be found in high school physics textbooks, we will not list them here.
When solving the exercises, we will simply recall and apply them.
Here we only mention the book by de Lange and Pierrus \cite[Section 7]{deLPie}
where one can find several interesting exercises with detailed solutions for two and three dimensional projectile motions.

The following is a list of some basic static concepts used in Section \ref{section_statics}.
A prismatic beam is a beam with a uniform cross-section throughout.
Reaction forces are the forces exerted by the supports on the beam to achieve equilibrium of a force system.
Two basic types of supports are used in this paper.
The first one is the pinned support, which can resist forces in any direction, but not rotation.
It allows the beam to rotate, but not to translate in any direction.
In this case the reaction force can be any force in the plane, i.e. its components can be different from zero in any direction.
The second one is the roller support which is free to rotate and translate along the surface on which the roller rests.
In this case, the reaction force is perpendicular to and away from the surface
 (in particular, its component parallel to the surface is zero).

The rotating effect of a force with respect to a (fixed) point is described by its moment of force,
 which is a scalar quantity in planar problems.
The convention is that the sign of the moment of force is positive if it rotates counterclockwise in the plane.
The magnitude of the moment of a force with respect to a point (point of rotation) is the product of the magnitude of the force
 and its lever arm, where the lever arm is the distance between the point of rotation and the line of action of the force.

To determine unknown reaction forces in a equilibrium situation of a rigid body in the plane, we can write equilibrium equations for the forces and the moments of forces.
One type of the equilibrium equations says that the sum of force vectors is zero.
In planar problems it means two scalar equations in two perpendicular directions (e.g. horizontal and vertical).
In planar problems, the other type of the equilibrium equations says that the sum of scalar moments of forces
 with respect to an arbitrary point of the plane is zero.
All in all, a rigid body in the plane is in equilibrium if the sum of the force vectors acting on it is zero and the sum of the corresponding
 scalar moments of forces with respect to an arbitrary point of the plane is zero.

\section{Alternative solution to part (ii) of Exercise \ref{Ex_proj_motion_1}}\label{appAltSol}

In this appendix, we determine the probabilities of the events $R$ and $S$ introduced
 in the solution of part (ii) of Exercise \ref{Ex_proj_motion_1} using the law of total expectation.
This is an alternative solution, since in the solution of part (ii) of Exercise \ref{Ex_proj_motion_1},
 we used the method of geometric probabilities.

First, we determine the probability $\PP(R)$.
Since $v$ is uniformly distributed on the interval $[5,15]$, its density function takes the form
 \[
  f_v(r) = \begin{cases}
               \frac{1}{10}  & \text{if \ $r\in[5,15]$,}\\
               0 & \text{otherwise.}
            \end{cases}
 \]
Using the law of total expectation (see \eqref{telj_varh_ertek_tetele}) and the independence of $h$ and $v$, we have that
 \begin{align*}
   \PP(R) & = \PP(d<10)
            = \PP\left(\sqrt{\frac{2h}{g}}\cdot v < 10\right)
            =  \int_{-\infty}^\infty \PP\left(\sqrt{\frac{2h}{g}}\cdot r < 10\right) f_v(r)\,\dd r\\
         & = \int_5^{15} \PP\left(\sqrt{\frac{2h}{g}}\cdot r < 10\right)\frac{1}{10}\,\dd r
           = \frac{1}{10} \int_5^{15} \PP\left( h < \frac{100g}{2} \cdot \frac{1}{r^2}\right)\,\dd r,
 \end{align*}
 where
 \[
 \PP\left( h < \frac{100g}{2} \cdot \frac{1}{r^2}\right)
  = \frac{1}{50}\min\left(50, \frac{50g}{r^2}\right)
  =\frac{1}{50} \cdot \frac{50g}{r^2} = \frac{g}{r^2},
  \qquad r\in[5,15],
 \]
 since $50g/r^2\leq 50g/25 = 19.62$.
Consequently,
 \[
  \PP(R) = \frac{1}{10} \int_5^{15} \frac{g}{r^2} \,\dd r = \frac{g}{10}\left(\frac{1}{5} - \frac{1}{15}\right)
         = \frac{g}{75} = 0.1308.
 \]

Next, we determine the probability $\PP(S)$.
Using the law of total expectation (see \eqref{telj_varh_ertek_tetele}) and the independence of $h$ and $v$, we have that
 \begin{align*}
 \PP(S)& = \PP\left( \frac{g}{2v^2}\cdot 10.5^2 < h-2\right)
        = \PP\left( h > 2+\frac{g}{2v^2}\cdot 10.5^2 \right) \\
       & = \frac{1}{10} \int_5^{15} \PP\left( h > 2+\frac{g}{2r^2}\cdot 10.5^2 \right)\,\dd r \\
       & = \frac{1}{10} \int_5^{15} \frac{1}{50} \max\Big( 50 - \Big(2+\frac{g}{2r^2}\cdot 10.5^2 \Big), 0\Big)\,\dd r,
 \end{align*}
 where $50 - \Big(2+\frac{g}{2r^2}\cdot 10.5^2 \Big)\geq 0$ for all $r\in[5,15]$, since
 it holds if and only if $r^2\geq 10.5^2g/96\approx 11.27$, which is true since $r\geq 5$.
Hence we get that
 \begin{align*}
   \PP(S) &= \frac{1}{500} \int_5^{15} \Big( 48 - \frac{g}{2r^2}\cdot 10.5^2 \Big)\,\dd r\\
          &=\frac{1}{500}\Big( 48(15-5) + \frac{10.5^2 g}{2}\Big(\frac{1}{15} - \frac{1}{5}\Big)  \Big)
           =\frac{48}{50} - \frac{110.25 g}{7500} \approx 0.8158.
 \end{align*}

Note that we calculated the same values for $\PP(R)$ and $\PP(S)$ as in the solution of part (ii) of Exercise \ref{Ex_proj_motion_1}, that was expected.

\section*{Acknowledgements}
We would like to thank G\'abor Dr\'otos, Janka V\'ertesi and Tam\'as V\'ertesi for their helpful comments and suggestions.
We are grateful to Ren\'ata Vas for helping us with the language.
We would like to thank the three referees for their comments that helped us to improve the paper.

\section*{Funding}
M\'aty\'as Barczy is supported by the Ministry of Innovation and Technology of Hungary
 from the National Research, Culture and Innovation Fund, project no.\ TKP2021-NVA-09.

\section*{Declaration of competing interest}
The authors declare that they have no known competing financial interests or personal relationships
 that could have appeared to influence the work presented in this paper.

\end{document}